\documentclass[pra,10pt,twocolumn,reprint,superscriptaddress,floatfix,showpacs]{revtex4-2}

\usepackage[utf8]{inputenc}
\usepackage{amsthm}
\usepackage[T1]{fontenc}     
\usepackage[british]{babel}  
\usepackage[sc,osf]{mathpazo}
\usepackage[scaled=0.86]{berasans}  
\usepackage[colorlinks=true, citecolor=blue, urlcolor=blue]{hyperref}  

\usepackage{graphicx} 
\usepackage{subfig}
\usepackage{bitset}
\usepackage[babel]{microtype}  
\usepackage{amsmath,amssymb,amsthm,bm,amsfonts,mathrsfs,bbm} 
\usepackage{xspace}  
\usepackage{pgfplots}
\usepackage{xcolor,colortbl}
\def\ba{\begin{equation}}
	\def\ea{\end{equation}}
\def\bea{\begin{eqnarray}}
	\def\eea{\end{eqnarray}}
\def\ben{\begin{equation*}}
	\def\een{\end{equation*}}
\def\bean{\begin{eqnarray*}}
	\def\eean{\end{eqnarray*}}
\def\bma{\begin{mathletters}}
	\def\ema{\end{mathletters}}
\def\bi{\begin{itemize}}
	\def\ei{\end{itemize}}

\newcommand{\be}{\begin{equation}}
	\newcommand{\ee}{\end{equation}}

\newcommand{\kommentar}[1]{}

\newcommand{\forget}[1]{}

\newtheorem{theorem}{Theorem}

\begin{document}
	
	\title{Network Nonlocality Without Entanglement Of All Sources}
	\author{Kaushiki Mukherjee}
	\email{kaushiki.wbes@gmail.com}
	\affiliation{Department of Mathematics, Government Girls' General Degree College, Ekbalpore, Kolkata-700023, India.}
	\author{Biswajit Paul}
	\email{biswajitbbkm01082019@gmail.com}
	\affiliation{Department of Mathematics, Balagarh Bijoy Krishna Mahavidyalaya, Hooghly, West Bengal, India.}
	
	\begin{abstract}
		Entanglement and nonlocality are two important nonclassical features of quantum correlations. Recently the study of quantum correlations in networks has undergone remarkable progress owing to technological development towards scalable quantum networks. However, compared to standard Bell scenario, manifestation of the interplay between these two aspects has received less attention in network scenarios featuring independent sources. In this work we have analyzed the relation between entanglement content of the sources and detectable non $n$-locality in two distinct network topologies(linear and star). We have studied the extremal violations of $n$-local inequalities(compatible with linear and star network) for any fixed amount of entanglement(in terms of concurrence) of the independent sources. It is observed that  each of the sources must be entangled for detecting non $n$-locality in linear network. However, the same is not true for star $n$-local network. Present analysis is revealing that entanglement of all the independent sources is not a necessity for generation of non $n$-local correlations in star topology. Characterization of sources in terms of minimum entanglement requirement for any fixed violation amount of the $n$-local inequalities is also provided. Interestingly, detection of non $n$-locality is ensured in the network if product of concurrence of all the sources involved exceeds $\frac{1}{2}.$

	\end{abstract}
	
	\maketitle
	
	
	\section{Introduction}\label{intro}
	The pioneering work by J.S.Bell in $1964$\cite{Bel} is considered to be one of the cornerstone in the development of quantum nonlocality\cite{brunrev}. Bell nonlocality refers to the phenomenon predicted by quantum mechanics and experimentally confirmed by Bell's theorem, which shows that particles can exhibit correlations that cannot be explained by any local hidden variable theory. Such form of correlations is usually known as Bell nonlocal correlations and associated experimental setup is commonly referred to as standard Bell experiment\cite{Bel}. Experimental detection of Bell nonlocal correlations relies upon Bell inequalities. Over years, several forms of Bell inequality have been framed compatible with different Bell scenarios. Quantum entanglement is necessary but not sufficient for Bell inequality violation and hence for detecting quantum nonlocality\cite{brunrev}. All pure states are Bell-nonlocal\cite{Gis,Pop}. However the same is not true for mixed quantum states\cite{Wer,Bar}. Multiple studies have been made in exploiting the inter relationship between mixed entanglement and Bell nonlocality\cite{brunrev,Nico,Hors,Bart,Augs}. In any such study analyzing a single quantum source, compatible with corresponding Bell scenario, is involved.\\
	\par In recent past, different striking features of nonlocality are detected in networks involving multiple quantum sources\cite{BGP,BRGP, Frit,Fritz, Woo, Hens, Cha, Tava, Tavak, Tavako, Chav, Ross, Tavakol}. Different experimental works based on quantum networks have inspired conceptual development of quantum nonlocality \cite{Sau,Car,Sun,Pod,Bau,And}. In any such network scenario, all the sources distribute quantum particles among the parties(nodes). Some of those nodes can perform joint quantum measurements thereby simulating stronger quantum correlations across the entire network\cite{Zuko}. Nonlocality(if any) of network correlations is referred to as \textit{network nonlocality}. In case the sources involved in the network are independent, network nonlocality is referred to as \textit{non} $n$-\textit{locality}($n$ denoting number of sources)\cite{Tavakol}. Non $n$-locality lead to phenomena that have no analogue in standard Bell scenarios. For instance, in contrast to Bell nonlocality, non $n$-nonlocality can be generated even when some of the parties perform a single measurement thereby lacking randomness(owing to choice of measurements). Also, non-classical correlations are created among nodes sharing no direct common past(ref self). Starting from the simplest case of non bilocality\cite{BGP, BRGP}, the theory of non $n$-locality has been studied rigorously\cite{Frit,Fritz, Woo, Hens, Cha, Tava, Tavak, Tavako, Chav, Ross, Tavakol,Kau0,Kau,Kau1,Kau2,Kau3,kau4,kau5,kau6,kau7}. Apart from foundational importance, it is becoming increasingly relevant to apply quantum network nonlocality in different information processing tasks. This is because of the rapid technological advancements towards scalable networks involving quantum sources\cite{Kimb,Weh,Koz}.\\
	\par Two basic components of studying non $n$-locality are its detection and its relationship with entanglement. One obvious way for detection is to prove the impossibility of constructing any $n$-local model\cite{BRGP}. But  such an approach is theoretically challenging owing to the non-convexity of the set of $n$-local correlations. An alternate approach is to exploit the violations of
	any $n$-local inequality compatible with the concerned network scenario. Any $n$-local inequality being correlators-based, the latter approach is practically feasible. Till date, two most commonly used $n$-local inequalities are $I_{n-linear}$ for linear chain\cite{Kau} and $I_{n-star}$(see sec.\ref{pre}) for star shaped network\cite{Tava}. For $n$$=$$2,$ $I_{n-linear}$ is the famous BRGP inequality  introduced in \cite{BRGP}. Interpreting relation between entanglement of sources and non $n$-locality detection is important for obvious reasons. However, study on such inter-relationship is limited so far. In \cite{Gisi}, the authors have shown that all pure entangled states violate BRGP inequality\cite{BRGP}. However this is not the case for mixed entanglement\cite{BRGP}. Present discussion will focus on exploiting the inter-relationship between non $n$-locality detection and entanglement content of the sources. In this context, some obvious queries need to be addressed:
	\begin{itemize}
		\item \textit{Is entanglement of all sources a necessity for violating any $n$-local inequality?} 
		\item \textit{Given any amount of violation of a $n$-local inequality, how much entanglement is present in the independent sources?}
		\item \textit{Given any fixed amount of source entanglement, how much amount of violation of a $n$-local inequality can be observed?}
		\item \textit{Given the fact that only entangled states are used in the network, how much entanglement of any source can ensure generation of non $n$-local correlations in the network?}
		
	\end{itemize}
	Present study intend to address these queries for both linear and non-linear(star) shaped $n$-local networks.\\
	\par  $I_{n-linear}$ and $I_{n-star}$ being the two most common $n$-local inequalities in linear and star shaped network scenarios, our study will focus on violation of these two inequalities. Concurrence will be considered as the measure of qubit entanglement for our purpose. For each of $I_{n-linear}$ and $I_{n-star},$ we prescribe two inequalities $\mathcal{I}^{(lin)}_1(\mathcal{I}^{(star)}_1)$ and $\mathcal{I}_2^{(lin)}(\mathcal{I}^{(star)}_2)$ for establishing the relation between violation of this inequality and the entanglement content(in terms of concurrence) of the sources in the linear(star) $n$-local network. \\
	\par From practical view point, $\mathcal{I}_1^{(lin)},\mathcal{I}_1^{(star)}$ will emerge as a device-independent entanglement detection criterion in $n$-local networks for linear and star configuration respectively. Our work will lead to quantification of the degree of violation of the $n$-local inequalities. Inter-relationship between concurrence and $n$-nonlocality will provide insight to the amount of entanglement required for simulating detectable non $n$-local correlations in different noisy environments.\\
	\par Rest of our work will be organized as follows: in sec.\ref{pre} we provide the preliminaries required for our discussion. In sec.\ref{linnow} and sec.\ref{starnow}, we provide results pertaining to relation between concurrence and violation of $I_{n-linear}$ and $I_{n-star}$ inequalities respectively. Using results obtained in preceding two sections, non $n$-locality obtained in the linear and star topologies are next compared in sec.\ref{starnow1}. Practical implications of our results are discussed in sec.\ref{prac} followed by concluding remarks in sec.\ref{conc}. 
	\section{Preliminaries}\label{pre}
	\subsection{Bloch Matrix Representation}
	Let $\rho$ denote an arbitrary two qubit state shared between the parties Alice and Bob. In terms of Bloch parameters $\rho $ can be expressed as follows:
	\begin{equation}\label{st4}
		\small{\rho}=\small{\frac{1}{4}(\mathbb{I}_{2\times2}+\vec{u}.\vec{\sigma}\otimes \mathbb{I}_2+\mathbb{I}_2\otimes \vec{v}.\vec{\sigma}+\sum_{k_1,k_2=1}^{3}r_{k_1k_2}\sigma_{k_1}\otimes\sigma_{k_2})},
	\end{equation}
	with $\vec{\sigma}$$=$$(\sigma_1,\sigma_2,\sigma_3), $ $\sigma_{j}$ representing Pauli operators along $3$ mutually perpendicular directions($j$$=$$1,2,3$). $\vec{u}$$=$$(u_1,u_2,u_3)$ and $\vec{v}$$=$$(v_1,v_2,v_3)$ represent the local bloch vectors($\vec{u},\vec{v}$$\in$$\mathbb{R}^3$) of  Alice and Bob respectively with $|\vec{u}|,|\vec{v}|$$\leq$$1$ and $(r_{i,j})_{3\times3}$ denoting the correlation tensor $\mathcal{R}$(a real matrix).
	The components of $\mathcal{R}$ are given by $r_{ij}$$=$$\textmd{Tr}[\rho\,\sigma_{i}\otimes\sigma_{j}].$ \\
	$\mathcal{R}$ can be diagonalized by subjecting to suitable local unitary operations\cite{Luo}:
	\begin{equation}\label{st41}
		\small{\rho}^{'}=\small{\frac{1}{4}(\mathbb{I}_{2\times2}+\vec{\mathfrak{a}}.\vec{\sigma}\otimes \mathbb{I}_2+\mathbb{I}_2\otimes \vec{\mathfrak{b}}.\vec{\sigma}+\sum_{j=1}^{3}w_{jj}\sigma_{j}\otimes\sigma_{j})},
	\end{equation}
	Here the correlation tensor is $\textbf{W}$$=$$\textmd{diag}(w_{11},w_{22},w_{33}).$ $w_{11},w_{22},w_{33}$ are the eigenvalues of $\sqrt{\mathcal{R}^T\mathcal{R}},$ i.e., singular values of $\mathcal{R}.$

	\subsection{Concurrence}
	Concurrence is a measure of entanglement\cite{Ben}. Consider any two-qubit state $\rho.$ Let $\mu_!,\mu_2,\mu_3,\mu_4$ denote the square of the eigen values of the matrix $\rho(\sigma_2\otimes \sigma_2)(\rho)^{T}(\sigma_2\otimes \sigma_2)$ arranged in decreasing order. Then concurrence($\mathcal{C}$) of $\rho$ is given by\cite{Woot}:
	\begin{equation}
		\mathcal{C}=\textmd{Max}[0,\sqrt{\mu_1}-\sum_{i=2}^{4}\mu_i]
	\end{equation}
	$\mathcal{C}$ ranges between $0$ and $1.$ In case of a separable state $\mathcal{C}$$=$$0$ whereas that for a maximally entangled state $\mathcal{C}$$=$$1.$
	\subsection{Quantum $n$-local Linear Network\cite{Kau}}\label{biloc}
\begin{center}
		\begin{figure}
			\includegraphics[width=3.4in]{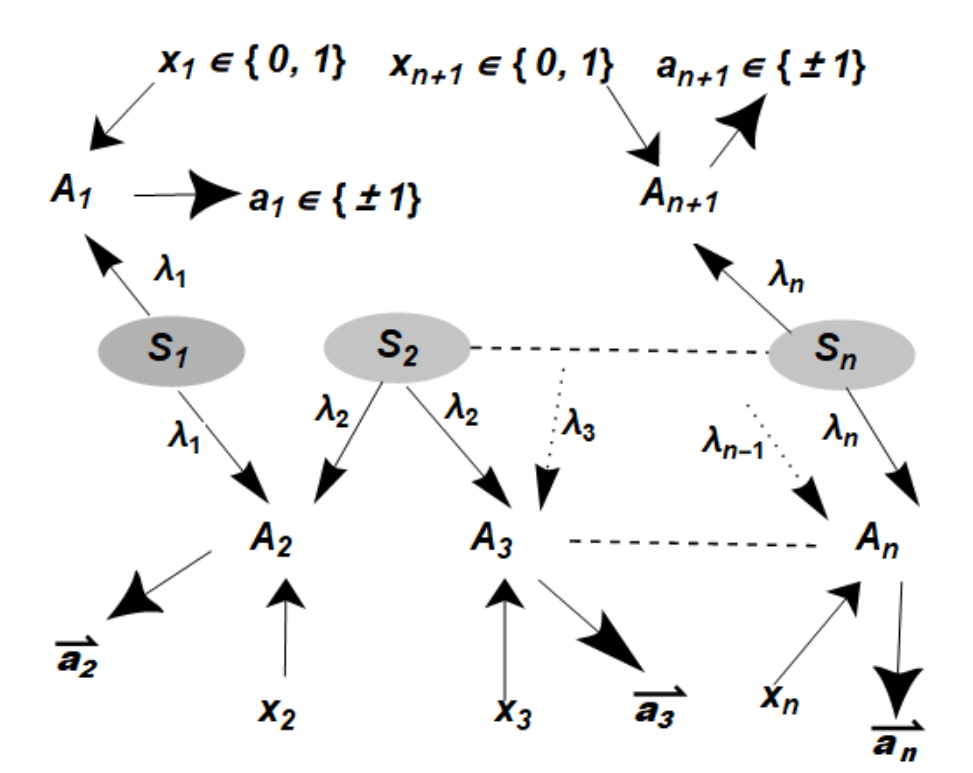} \\
			\caption{\emph{Schematic Diagram of linear $n$-local network.}}
			\label{slinear}
		\end{figure}
	\end{center}

	$n$ independent quantum sources and $n+1$ parties are arranged in a linear pattern (see Fig.\ref{slinear}). $\forall i$$=$$1,2,...,n,$ let $\mathcal{S}_i$ distribute a two qubit state $\rho_{i}.$  Let us denote the overall state in the network as $\rho_{in}$:
	\begin{equation}\label{tr4}
		\rho_{in}=\otimes_{j=1}^n \rho_{j}.
	\end{equation}
	Here $A_2,A_3,...,A_{n}$ are the intermediate parties and $A_1,A_{n+1}$ are the extreme parties. Let each of $A_i(i$$=$$2,...,n)$ perform a single measurement with $4$ outcomes\cite{BRGP,Kau} on the state of two qubits received from $\mathcal{S}_{i-1},\mathcal{S}_{i}$. Each of the two extreme parties performs any one of two projective measurements: $A_1$ and $A_{n+1}$ perform $x_1$ and $x_{n+1}$ respectively($x_1,x_{n+1}$$\in$$\{0,1$\}). \\
	$n+1$-partite correlations are $n$-local if those satisfy the following decomposition \cite{Kau}:
	\begin{eqnarray}\label{tr1}
		P(a_1,\vec{a}_2,...,\vec{a}_{n},a_{n+1})=	\int...\int_{\Lambda_1,...\Lambda_n}d\lambda_1,...d\lambda_n \sigma(\lambda_1,...\lambda_n) \textbf{P},&&\nonumber\\
		\textbf{P}= P(a_1|x_1,\lambda_1) P(a_{n+1}|x_{n+1},\lambda_n)\Pi_{2,...,n}P(a_{i1},a_{i2}|\lambda_{i-1},\lambda_i)&&\nonumber\\
		\textmd{with } \vec{a}_i=(a_{i1},a_{i2}),\forall i \quad\quad&&
	\end{eqnarray}
	along with the $n$-local constraint:
	\begin{equation}\label{tr2}
		\sigma(\lambda_1,...\lambda_n)=\Pi_{i=1}^n\sigma_i(\lambda_i)
	\end{equation}
	$\sigma(\lambda_1,...\lambda_n)$ denoting the probability density function of local hidden variables $\lambda_1,...,\lambda_n$ and $\sigma_i(\lambda_i)$ is defined analogously. $\vec{a}_i$$=$$(a_{i1},a_{i2})$$\in$$\{(0,0),(0,1),(1,0),(1,1)\}$ denote the $4$ outcomes of party $A_i(i$$=$$2,3,...,n).$ $a_1,a_{n+1}$$\in$$\{0,1\}$ denote binary outputs of $A_1$ and $A_{n+1}$ respectively.
	So, any set of $n+1$-partite correlations not satisfying both of these restrictions (Eqs.(\ref{tr1},\ref{tr2})), are termed as non $n$-local.
	The existing $n$-local inequality($I_{n-linear}$,say) is given by\cite{Kau}:
	\begin{eqnarray}\label{ineqb}
		&& \sqrt{|\mathcal{I}_n|}+\sqrt{|J_n|}\leq  1,\,  \textmd{where}\\
		&& I_n=\frac{1}{4}\sum_{x_1,x_{n+1}}\langle D_{1,x_1}D_2^0D_3^0...D_{n}^0D_{n+1,x_{n+1}}\rangle\nonumber\\
		&&   J_n= \frac{1}{4}\sum_{x_1,x_{n+1}}(-1)^{x_1+x_{n+1}}\langle \small{ D_{1,x_1}D_2^1...D_{n}^1D_{n+1,x_{n+1}}}\rangle\,\,\textmd{with} \nonumber\\
		&&   \langle D_{1,x_1}D_2^i...D_{n}^iD_{n+1,x_{n+1}}\rangle = \sum_{\mathcal{D}_1}(-1)^{\textbf{a}_1+\textbf{a}_{n+1}+\sum_{j=2}^{n}\textbf{a}_{j(i+1)}}Q_1,\nonumber\\
		&& \textmd{\small{where}}\,Q_1=\small{p(\textbf{a}_1,
			\vec{a}_2,...,\vec{a}_{n},
			\textbf{a}_{n+1}|x_1,x_{n+1})},\, i=0,1\nonumber\\
		&& \mathcal{D}_1=\{\textbf{a}_1,\textbf{a}_{21},\textbf{a}_{22},...,
		\textbf{a}_{n1},\textbf{a}_{n2},\textbf{a}_{n+1}\}\nonumber
	\end{eqnarray}
	Violation of $I_{n-linear}$(Eq.(\ref{ineqb})) guarantees that non $n$-locality of corresponding correlations. No definite conclusion can be given in case $I_{n-linear}$ is satisfied.\\
	Now, $\forall i,$ let $\mathcal{S}_i$ distribute any two qubit state $\rho_{i}$ (Eq.\ref{st41}). Each of $A_2,...,A_n$ perform Bell basis measurement(BSM). $A_1$ and $A_{n+1}$ each perform qubit projective measurement in any one of two directions. 
	
	In such measurement context, the upper bound of Eq.(\ref{ineqb}) is given by\cite{Gisi}:
	\begin{equation}\label{boundlin}
		B_{n-linear}=\sqrt{\Pi_{j=1}^n E_{j1}+\Pi_{j=1}^n E_{j2}}
	\end{equation}
	
		with $E_{i1},E_{i2},E_{i3}$$\in$$\{w_{i1},w_{i2},w_{i3}\}$ such that    $E_{i1}$$\geq$$E_{i2}$$\geq$$E_{i3}$ denote the ordered eigen values of $\sqrt{(W_i)^TW_i}, \forall i$\\
	
 $I_{n-linear}$ is thus violated if:
	\begin{equation}\label{up11}
		B_{n-linear}>1
	\end{equation}
	So, if Eq.(\ref{up11}) is not satisfied, then corresponding tripartite correlations are $n$-local up to the $n$-local inequality $I_{n-linear}$(Eq.(\ref{ineqb})).
	\subsection{Quantum Star Network\cite{Tava}}\label{star1}
	A star-shaped network, or simply a star network is a non-linear $n$-local network. It consists of one intermediate party (central party) $A_1$ and $n$ extreme parties $A_2,...,A_{n+1}$ (see Fig.\ref{star1}). $A_1$ receives $n$ qubits(one from each source $\mathcal{S}_i$). Each of remaining parties receives single qubit. \\
	Each of the $n$$+$$1$ parties performs any one of two projective measurements. $n+1$-partite correlations are said to be $n$-local if those simultaneously satisfy Eq.(\ref{tr2}) and the following decomposition \cite{Tava}:
	\begin{eqnarray}\label{tr3}
		P(\vec{a}_1,a_2,...,a_{n+1})=	\int...\int_{\Lambda_1,...\Lambda_n}d\lambda_1,...d\lambda_n \sigma(\lambda_1,...\lambda_n) \textbf{Q},&&\nonumber\\
		\small{where}\,\textbf{Q}= \Pi_{i=2}^{n+1}P(a_i|x_i,\lambda_{i-1})P(a_1|x_1
		\lambda_1,...,\lambda_n)&&\nonumber\\
	\end{eqnarray} 
	$n+1$-partite correlations not satisfying both of these restrictions (Eqs.(\ref{tr2},\ref{tr3})), are termed as non $n$-local. 
	The existing $n$-local inequality for star network is given by\cite{Tava}:
	\begin{eqnarray}\label{ineqs}
	 \sum_{i=1}^{2}|J_{i}|^{\frac{1}{n}}\leq  2^{n-2},\,  \textmd{where}&&\\
		 J_i=\frac{1}{2^n}\sum_{x_1,...,x_{n+1}}(-1)^{h_i\sum_{j=1}^nx_j}&&\nonumber\\
		 \langle A_{x_1}^{(1)}A_{x_2}^{(2)}...A_{x_{n+1}}^{(n+1)}\rangle&&\nonumber\\
	\textmd{where }	\langle A_{(1)}^{(1)}A_{x_2}^{(2)}...A_{x_{n+1}}^{(n+1)}\rangle=&&\nonumber\\
		\sum_{a_1,...,a_{n+1}}(-1)^{\sum_{j=1}^{n+1}a_j}P(a_1,...,a_{n+1}|x_1,...,x_{n+1})&&\nonumber\\
		h_1=0\,\textmd{and }h_2=1&&\nonumber\\
	\end{eqnarray}
Let each of the sources distribute an arbitrary two qubit state (Eq.\ref{st41}). Let each of $A_1,A_2,...,A_{n+1}$ perform projective measurements. In such scenario, the upper bound of Eq.(\ref{ineqs}) is given by\cite{And}: 
	\begin{equation}\label{boundstar}
		B_{n-star}=	\sqrt{(\Pi_{j=1}^n  E_{j1})^{\frac{2}{n}}+(\Pi_{j=1}^nE_{j2})^{\frac{2}{n}}}.
	\end{equation}
	
	$n$-local inequality (Eq.(\ref{ineqs})) is thus violated if:
	\begin{equation}\label{up211}
		B_{n-star}>1.
	\end{equation}
	Hence, if Eq.(\ref{up211}) is violated, then corresponding correlations turn out to be $n$-local as per the $n$-local inequality (Eq.(\ref{ineqs})). 
		\begin{center}
		\begin{figure}
			\includegraphics[width=3.4in]{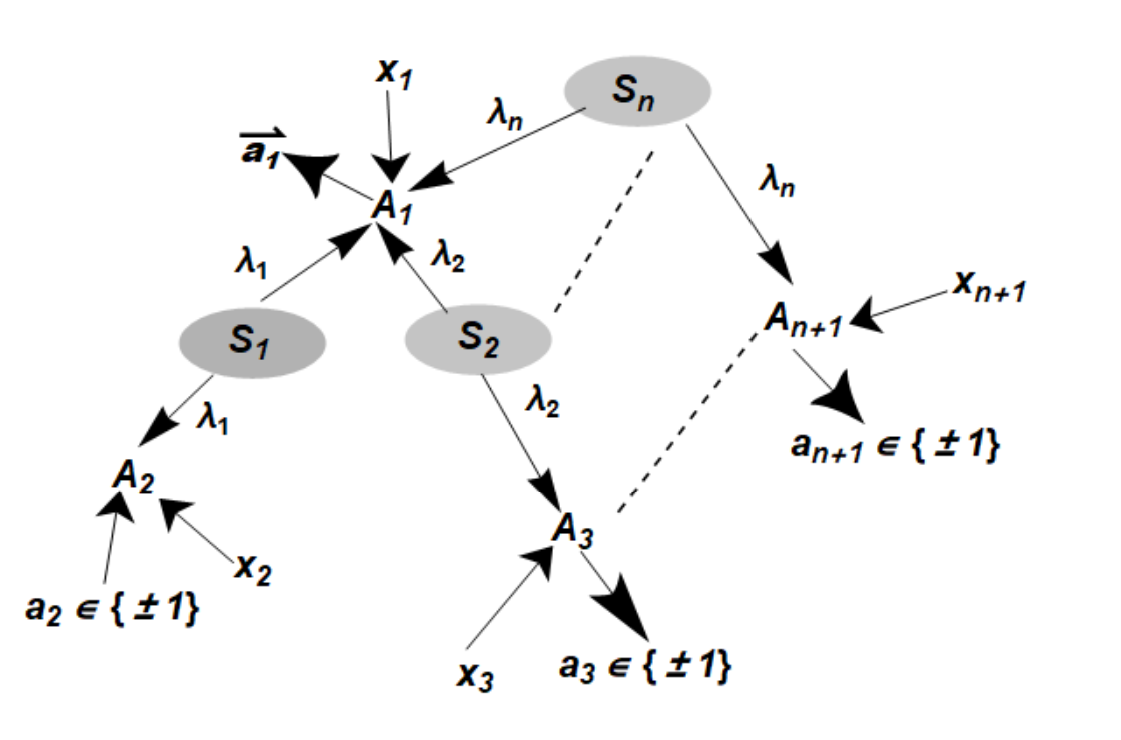} \\
			\caption{\emph{Schematic Diagram of $n$-local star network.}}
			\label{star1}
		\end{figure}
	\end{center}
	\subsection{X Class of Two-Qubit States }\label{xstates}
	X Class of Two-Qubit States $\chi$ is given in the following form\cite{Tyo}: 
	\begin{equation}\label{xst}
		\chi=	\left[ {\begin{array}{cccc}
				x_{1}&0&0&y_{1}\\
				0&x_{2}&y_{2}&0\\
				0&y_{2}&x_{3}&0\\
				y_{1}&0&0&x_{4}\\		
		\end{array} } \right]
	\end{equation}
	For physical state criteria\cite{Tyo}, one requires $x_i$$\geq$$0,$ $x_1$$+$$x_2$$+$$x_3$$+$$x_4$$=$$1,$ $y_1^2$$\leq$$x_4x_1$ and $y_2^2$$\leq$$x_3x_2.$\\
	\par This family of states includes many well-known class of states such as Bell diagonal family\cite{Horo}, Werner states\cite{Wern}. This class is available in several experimental works\cite{Chi,Ldi,Nap,Pra,Sbo,Jwa}. In \cite{Maz} claims possibility of converting any two-qubit state state into X state by passing the state through a noisy channel. In \cite{Tuj}, the authors showed invariance of some forms of X states under general decoherence. This class has been used for exploiting sudden death of entanglement in \cite{Vin}. 
	\section{Relation Between $I_{n-linear}$ and $\mathcal{C}$}\label{linnow}
	Before we start our discussions, let us first specify the concerned linear network scenario.
	\subsection{Linear $n$-Local Scenario}
	Consider a network with $n$ independent sources and $n+1$ parties(see Fig.\ref{slinear}). Each of $A_2,A_3,...,A_n$ performs a single measurement: Bell basis measurement(BSM). Extreme parties $A_1$ and $A_{n+1}$ each performs projective measurement in two arbitrary directions. As already specified, correlations generated in such measurement context violate $I_{n-linear}$(Eq.\ref{ineqb}) if Eq.(\ref{up11}) is satisfied.\\
	\par For above network scenario, violation of $I_{n-linear}$ acts as a sufficient detection criterion for non $n$-locality. So, we next proceed to derive relation between the upper bound($B_{n-linear}$) of $I_{n-linear}$ and entanglement content of the sources used in the network. For that we subsequently derive two inequalities: $I_1^{(lin)}$ and $I_2^{(lin)}$(say). $I_1^{(lin)}$ and $I_2^{(lin)}$ provides an upper and a lower bound on $B_{n-linear}$ respectively in terms of the source entanglement contents.
	
	\subsection{Derivation of $I_1^{(lin)}$}\label{i1-lin}
	Let us first consider the following theorem relating $\mathcal{B}_{n-linear}$ and $\mathcal{C}_i$$(i$$=$$1,2,...,n$):
	\begin{theorem}\label{theo1}
	 In any $n$-local linear network with $\mathcal{S}_i$ distributing arbitrary two-qubit state $\rho_i$ with concurrence $\mathcal{C}_i$$(i$$=$$1,2,...,n$), $\mathcal{B}_{n-linear}$ and $\mathcal{C}_i$$(i$$=$$1,2,...,n$) satisfy the following relation:
	\begin{equation}\label{up1}
		\mathcal{B}_{n-linear}\leq \sqrt{1+\Pi_{i=1}^n\mathcal{C}_i}
	\end{equation} 
	\end{theorem}
	\textit{Proof:} See Appendix.A\\
	We refer to Eq.(\ref{up1}) as the inequality \textbf{$I_1^{(lin)}$}.
	An interesting result is implied by $I_1^{(lin)}$ which we put below as a corollary.\\
	\textit{Corollary.1:}\label{cor1} For detecting non $n$-locality via violation of $I_{n-linear},$ each of the sources $\mathcal{S}_1,\mathcal{S}_2,...,\mathcal{S}_n$ must distribute two-qubit entangled state in the network. \\
	Above corollary points out the impossibility of detecting non $n$-locality in the linear $n$-local network even if one of the sources distribute separable two-qubit state.
	\subsubsection{Upper Bound on Maximum Violation of $I_{n-linear}$} Inequality $I_1^{(lin)}$ provides an upper bound on the maximum(with respect to measurement parameters) possible violation of $I_{n-linear}$ in terms of entanglement content. Maximum value(with respect to all two-qubit states) of $\mathcal{B}_{n-linear}$ is $\sqrt{2}.$ That value can be obtained from Eq.(\ref{up1}) when each of $\mathcal{S}_1,\mathcal{S}_2,...,\mathcal{S}_n$ distribute maximally entangled qubits.\\
	Let $S_1$$=$$(\rho_1,\rho_2,...,\rho_n)$	and $S_2$$=$$(\rho_1^{'},\rho_2^{'},...,\rho_n^{'})$ represent two different sets of states such that for all $i$$=$$1,2,...,n,$ both $\rho_i$ and $\rho_i^{'}$ have same concurrence $\mathcal{C}_i.$ Above theorem ensures that amount of violation of $I_{n-linear}$ for none of $S_1$ and $S_2$ can exceed the limit of $\sqrt{1+\Pi_{i=1}^n\mathcal{C}_i}.$ \\
	\par Conversely, let $V$ denote the amount of violation of $I_{n-linear}.$ Then $B_{n-linear}$$=$$1+V.$ Also consider the quantity  $\mathcal{K}$$=$$\Pi_{i=1}^n\mathcal{C}_i.$ Clearly, $\mathcal{K}$ is a function of the entanglement content of all the sources used in the network. Let any fixed amount of violation $V$ be obtained in the linear network. In that case, $I_1^{(lin)}$ implies:
	\begin{equation}\label{fv1}
		\mathcal{K}\geq(1+V)^2-1.
	\end{equation}
	Eq.(\ref{fv1}) clearly points out the minimum amount of source entanglement required for $V$ amount of violation.
	Expression in R.H.S. of Eq.(\ref{up1}) clearly points out that equality in $I_1^{(lin)}$ does not hold for many sets of $n$ two-qubit states. For any set of entangled states not violating $I_{n-linear},$ Eq.(\ref{up1}) will be a strict inequality. Also, $I_1^{(lin)}$ may not be tight for entangled states violating $I_{n-linear}.$\\
	For example, consider the following class of two-qubit entangled states:
	\begin{eqnarray}\label{horodeck}
	\rho&=&p |\psi^{+}\rangle\langle\psi^{+}|+(1-p)|00\rangle\langle00|,\,\small{\textmd{where}}\\
	|\psi^{+}\rangle&=&\frac{|01\rangle+|10\rangle}{\sqrt{2}}\,\,p\in [0,1]
	\end{eqnarray}
	Concurrence of this class is given by $p.$
	Let each of $\rho_i$($i$$=$$1,...,n$) is a member of this class(Eq.(\ref{horodeck})) with parameter $p_i$$\in$$(\frac{1}{3},1)$. Using Eq.(\ref{up11}), it is clear that $I_{n-linear}$ is violated if 
	\begin{equation}\label{hor2}
		B_{n-linear}=\sqrt{2\Pi_{i=1}^n p_i}>1
	\end{equation}
	For this set of states, R.H.S. of $I_1^{(lin)}$ is given by $\sqrt{1+\Pi_{i=1}^n p_i}.$ By Eq.(\ref{hor2}), $B_{n-linear}$$<$$\sqrt{1+\Pi_{i=1}^n p_i}.$ Hence, $I_1^{(lin)}$ is a strict inequality here. \\
	In this context, it now becomes imperative to find out states for which equality holds in Eq.(\ref{up1}). 
	\subsubsection{Examples of states for Which $I_1^{(lin)}$Eq.(\ref{up1}) Is Tight}
	Following theorem provides examples of two-qubit states for which the upper bound provided by Theorem.\ref{theo1} is satisfied.
	\begin{theorem}\label{theo2}
	 Let $S$$=$$(\rho_1\,\rho_2,...,\rho_n)$ denote any set of $n$ entangled two-qubit states. Equality will hold in $I_1^{(lin)}$ if:\\
	\begin{itemize}
		\item $\rho_i$ is a pure state $\forall i$
		\item  $\rho_i$ is a Rank-$2$ Bell-Diagonal state $\forall i$
		\item $S$ contain both of the above two types of states. 
	\end{itemize}
	\end{theorem}
	\textit{Proof:} See Appendix.B\\
	$B_{n-linear}$$=$$\sqrt{1+\Pi_{i=1}^n\mathcal{C}_i}$ in any $n$-local network involving the categories of two-qubit states as specified in Theorem.\ref{theo2}. So for any fixed violation amount($V$) of $I_{n-linear},$ Theorem.\ref{theo2} prescribes the condition to be satisfied by the entanglement content of the states involved in the network:
	\begin{equation}\label{up2}
		\mathcal{K}=(1+V)^2-1.
	\end{equation}
	In case all the states are identical, Eq.(\ref{up2}) points out the minimum entanglement content($\mathcal{C}$) required for $V$ amount of violation:
	\begin{equation}\label{up3}
		\mathcal{C}=((1+V)^2-1)^{\frac{1}{n}}
	\end{equation}
	\subsection{Derivation of $I_2^{(lin)}$ }\label{i2-lin}
\begin{center}
	\begin{figure}[!ht]
		\begin{tabular}{c}
			\subfloat[ ]{\includegraphics[trim = 0mm 0mm 0mm 0mm,clip,scale=0.6]{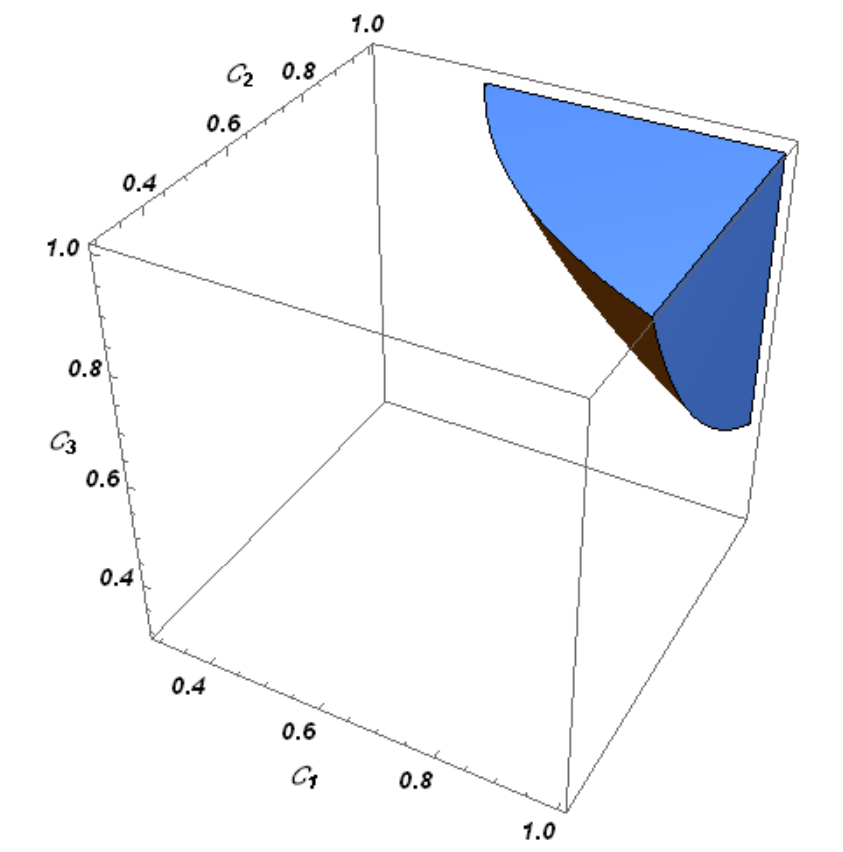}}\\
			\subfloat[]{\includegraphics[trim = 0mm 0mm 0mm 0mm,clip,scale=0.6]{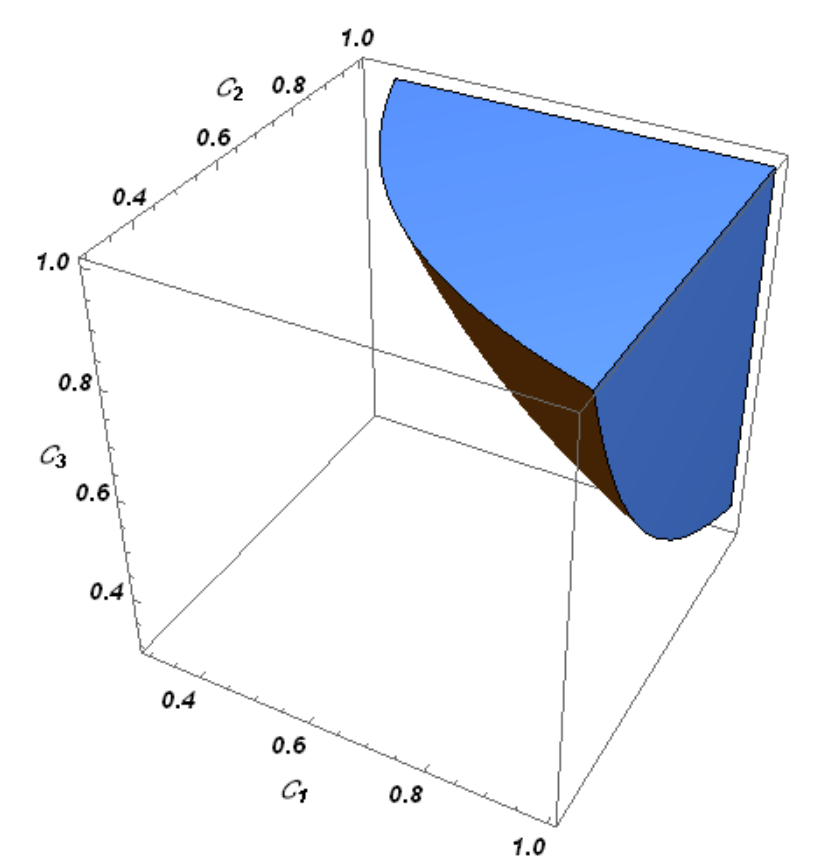}}\\
		\end{tabular}
		\caption{\emph{Sub-fig.(a) and (b) gives range of concurrences $\mathcal{C}_1,\mathcal{C}_2,\mathcal{C}_3$ of arbitrary Bell-Diagonal states for which non trilocality is detected in linear network and star trilocal network respectively. Comparison of the two sub-figures points out that for some values of $\mathcal{C}_1,\mathcal{C}_2$ and $\mathcal{C}_3$ non trilocality can only be detected for star topology.}}
		\label{figstarlower}
	\end{figure}
\end{center}

	Here we intend to characterize the lower bound of $B_{n-linear}.$ The theorem below provides the same for Bell diagonal class of states\cite{Horo}.
	\begin{theorem}\label{theo3}
	 In any $n$-local linear network where $\mathcal{S}_i$ distributes arbitrary Bell-diagonal state $\rho_i$ with concurrence $\mathcal{C}_i$$(i$$=$$1,2,...,n$), $\mathcal{B}_{n-linear}$ is lower bounded by a function of $\mathcal{C}_i$$(i$$=$$1,2,...,n$) as follows:
	\begin{equation}\label{up4}
		\mathcal{B}_{n-linear}\geq \sqrt{2\Pi_{i=1}^n\mathcal{C}_i}
	\end{equation}
	\end{theorem}
	\textit{Proof:} See Appendix.C\\ 
	Eq.(\ref{up4}) will be referred to as \textbf{$I_2^{(lin)}$} for further discussion in this section. $I_2^{(lin)}$ clearly points out that non $n$-locality can be detected in a linear network using Bell diagonal states satisfying the relation(see sub-fig.(a) in Fig.\ref{figstarlower} for trilocal network):
	\begin{equation}\label{up5}
		\Pi_{i=1}^n\mathcal{C}_i>\frac{1}{2}.
	\end{equation} 
	Alternatively, Eq.(\ref{up5}) can also be interpreted as a criterion quantifying the entanglement content of Bell-diagonal states generating non $n$-local network correlations. \\
	
\subsubsection{Applying $I_2^{(lin)}$ Beyond Bell-Diagonal Class} 
\begin{center}
		\begin{figure}[!ht]
				\begin{tabular}{c}
				\subfloat[ ]{\includegraphics[trim = 0mm 0mm 0mm 0mm,clip,scale=0.5]{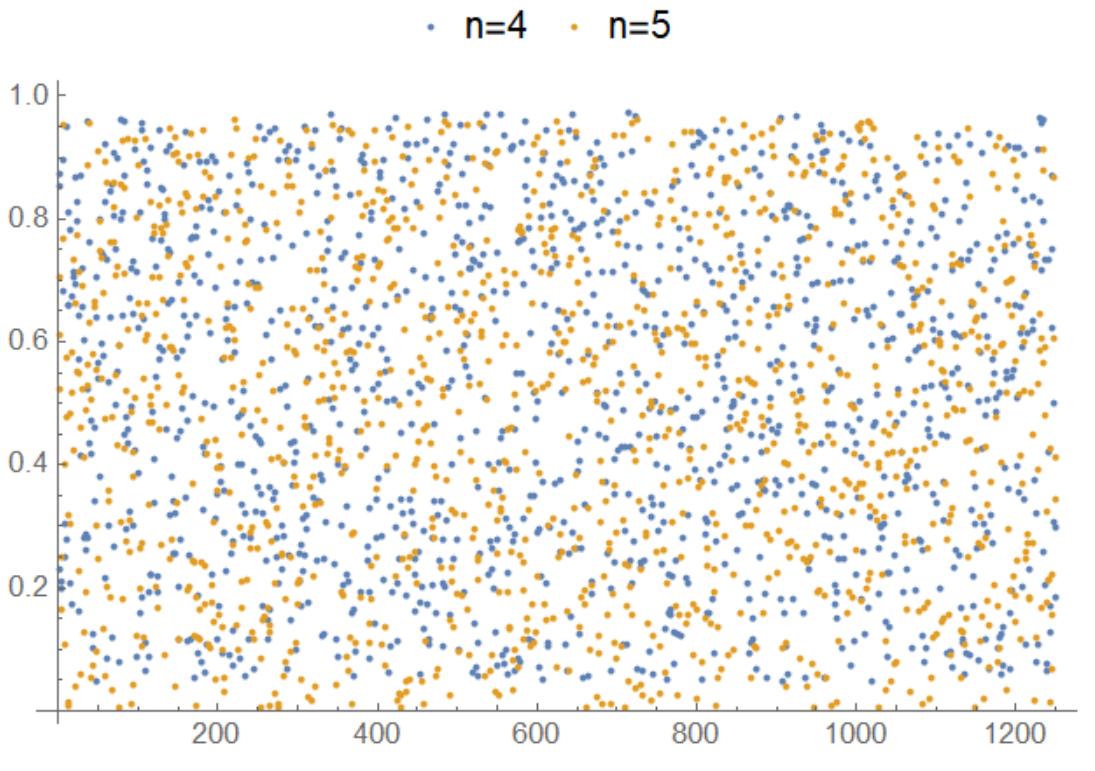}}\\
				\subfloat[]{\includegraphics[trim = 0mm 0mm 0mm 0mm,clip,scale=0.6]{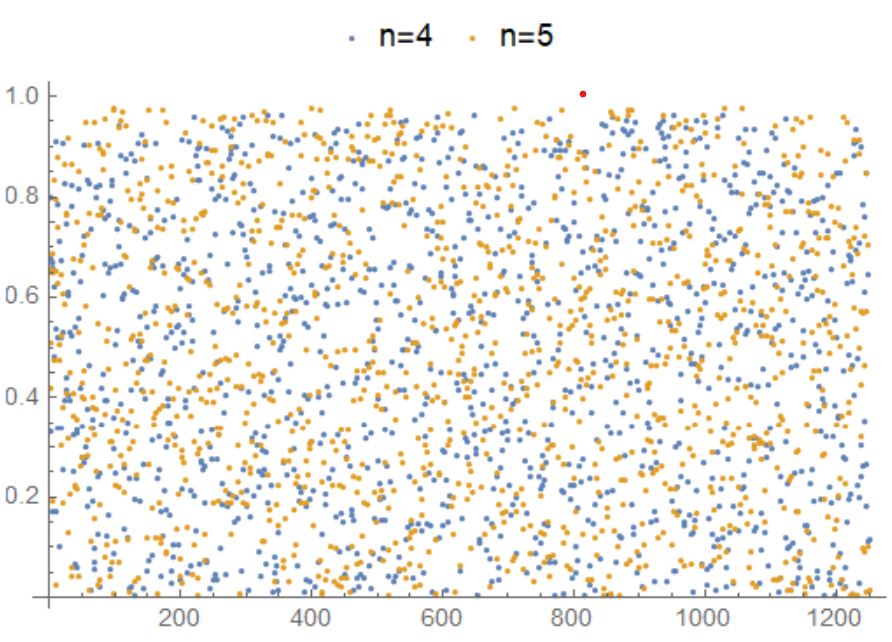}}\\
			\end{tabular}
			\caption{\emph{Considering $n$$=$$4,5$(differently colored points), $\mathcal{B}_{n-linear}$$-$$ \sqrt{2\Pi_{i=1}^n\mathcal{C}_i}$(sub-fig.(a)) and $\mathcal{B}_{n-star}$$-$$ \sqrt{2}\Pi_{i=1}^n\mathcal{C}_i^{\frac{1}{n}}$(sub-fig.(b)) is plotted along vertical axis for $>$$10^3$ sets of random two-qubit states. $I_2^{(lin)}$ and $I_2^{(star)}$ are satisfied for every plotted instance as region containing the points include only positive direction of vertical axis in both sub-figures. }}
			\label{linrand}
		\end{figure}
\end{center}

	As per Theorem.\ref{theo3}, $I_2^{(lin)}$ holds for Bell-Diagonal class of states. Next we will provide examples of two qubit states lying outside this family for which $I_2^{(lin)}$ holds. For that we consider a subclass of two-qubit $X$ states(Eq.(\ref{xst})). Any state from that subclass will be given by Eq.(\ref{xst}) for $x_4$$=$$y_1$$=$$0.$\\
	Let $\rho_i(i$$=$$1,2,...,n)$ be any member from this subclass:
	\begin{equation}\label{xst2}
		\rho_i=\chi_i=	\left[ {\begin{array}{cccc}
				x_{i1}&0&0&0\\
				0&x_{i2}&y_{i2}&0\\
				0&y_{i2}&x_{i3}&0\\
				0&0&0&0\\		
		\end{array} } \right]
	\end{equation}
	Concurrence of $\rho_i$ is given by: $\mathcal{C}_i$$=$$2y_{i2}.$ \\
	Correlation tensor of $\rho_i$ turns out to be:\textmd{diag}$(2y_{i2},2y_{i2},x_{i1}-x_{i2}-x_{i3}).$ \\
	Three cases may arise:
	\begin{enumerate}
		\item $2y_{i2}$$\geq$$x_{i1}-x_{i2}-x_{i3},$ $\forall$ $i$$=$$1,2,...,n$
		\item $2y_{i2}$$\leq$$x_{i1}-x_{i2}-x_{i3},$ $\forall$ $i$$=$$1,2,...,n$
		\item $2y_{i2}$$\geq$$x_{i1}-x_{i2}-x_{i3}$  $\forall  i$$\in$$H_1,$ \small{and}\\
		$2y_{i2}$$\leq$$x_{i1}-x_{i2}-x_{i3}$  $\forall i$$\in$$H_2,$ \small{such that} \\
		$H_1$$\cup$$H_2$$=$$\{1,2,...,n\}$ and $H_1$$\cap$$H_2$$=$$\Phi.$
	\end{enumerate}
	\textbf{Case.1:} $B_{n-linear}(Eq.(\ref{boundlin}))$ turns out to be:
	\begin{eqnarray}\label{xst3}
		B_{n-linear}&=&\sqrt{2\Pi_{i=1}^n(2y_{i2})}\nonumber\\
		&=&\sqrt{2\Pi_{i=1}^n\mathcal{C}_i}
	\end{eqnarray}
	Clearly $I_1^{(lin)}$ is satisfied here.\\
	\textbf{Case.2:} 
	Here $B_{n-linear}$ is given by:
	\begin{eqnarray}\label{xst4}
		B_{n-linear}&=&\sqrt{\Pi_{i=1}^n(x_{i1}-x_{i2}-x_{i3})+\Pi_{i=1}^n(2y_{i2})}\nonumber\\
		&=&\sqrt{\Pi_{i=1}^n(x_{i1}-x_{i2}-x_{i3})+\Pi_{i=1}^n\mathcal{C}_i}\nonumber\\
		&\geq& \sqrt{\Pi_{i=1}^n\mathcal{C}_i+\Pi_{i=1}^n\mathcal{C}_i}\,\small{\textmd{by condition of the case}}\nonumber\\
		&=&\sqrt{2\Pi_{i=1}^n\mathcal{C}_i }
	\end{eqnarray}
	\textbf{Case.3:} In this last possible case, $B_{n-linear}$ becomes:
	\begin{eqnarray}\label{xst5}
		B_{n-linear}&=&\sqrt{\Pi_{i\in H_1,j\in H_2}(x_{i1}-x_{i2}-x_{i3})(2y_{j2})+\Pi_{i=1}^n(2y_{i2})}\nonumber\\
		&=&\sqrt{\Pi_{i\in H_1}(x_{i1}-x_{i2}-x_{i3})\Pi_{i\in H_2}\mathcal{C}_i+\Pi_{i=1}^n\mathcal{C}_i}\nonumber\\
		&\geq& \sqrt{(\Pi_{i\in H_1}\mathcal{C}_i)(\Pi_{j\in H_2}\mathcal{C}_j)+\Pi_{i=1}^n\mathcal{C}_i}\nonumber\\
		&&\quad\quad \small{\textmd{by condition of the case}}\nonumber\\
		&=&\sqrt{2\Pi_{i=1}^n\mathcal{C}_i }
	\end{eqnarray}
	Eqs.(\ref{xst3},\ref{xst4},\ref{xst5}) imply that $I_1^{(lin)}$ is satisfied in each of the three cases.\\
	Besides, $I_2^{(lin)}$ has been tested over several random sets of two-qubit states and the inequality holds for all such numerical instances. To be precise, the difference $\mathcal{B}_{n-linear}$$-$$ \sqrt{2\Pi_{i=1}^n\mathcal{C}_i}$ is calculated for some random set of two-qubit states. For all such instances this difference turns out to be non-negative(see sub-fig.(a) in Fig.\ref{linrand}). Based on our numerical findings, we give the following conjecture:\\
	\textit{\textbf{Conjecture.1:}$I_2^{lin}$ holds for any set of two-qubit states.}
	
\subsubsection{Tightness of $I_2^{(lin)}$(Eq.(\ref{up4}))}
	Consider each of the $n$ sources $\mathcal{S}_i$ distributing any one of the four Bell states.\\
	For these states $B_{n-linear}$$=$$\sqrt{2}.$ Again concurrence of each state is $1.$ Hence, $I_2$ turns out to be an equality for Bell states.\\
	Another instance of tightness is provided by Eq.(\ref{xst3}) for the subclass of $X$ states(Eq.(\ref{xst2})) satisfying the condition $2y_{i2}$$\geq$$x_{i1}-x_{i2}-x_{i3},$ $\forall$ $i$$=$$1,2,...,n.$ 
	\subsubsection{Examples of states not saturating $I_2^{(lin)}$}
	There exist many entangled states for which $I_2^{(lin)}$ turns out to be a strict inequality. One such example is provided by the class given by Eq.(\ref{horodeck}) specified by the noise parameter $p$ lying in the range $[0,\frac{1}{3}].$ Let each of $\rho_1,...,\rho_n$ belong to this subclass. Then $B_{n-linear}$$=$$\sqrt{\Pi_{i=1}^n p_i+\Pi_{i=1}^n(1-2p_i)}.$ Here $I_2^{(lin)}$ is not saturated as:
	\begin{eqnarray}
		\sqrt{\Pi_{i=1}^n p_i+\Pi_{i=1}^n(1-2p_i)}&>&\sqrt{2\Pi_{i=1}^np_i}\nonumber\\
		\textmd{as } p_i&<&1-2p_i\,\forall\, i=1,2,...,n\nonumber
	\end{eqnarray}
	\section{Relation Between $I_{n-star}$ and $\mathcal{C}$}\label{starnow}
	\subsection{Star $n$-Local Scenario}
	Consider a star network with $n$ independent sources and $n+1$ parties(see Fig.\ref{star1}). Each of $A_1,A_2,...,A_{n+1}$ performs projective measurement in two arbitrary directions\cite{And}. As discussed in sec.\ref{pre}, correlations generated in such measurement context violate $I_{n-star}$(Eq.\ref{ineqs}) if Eq.(\ref{up211}) is satisfied. \\
	\par In star network scenario, violation of $I_{n-star}$ is sufficient for detecting non $n$-locality. We next proceed to establish relation between $B_{n-star}$ and entanglement content of the sources used in the network. As in linear network scenario, here also we derive $I_1^{(star)}$ and $I_2^{(star)}.$	
	\subsection{Derivation of $I_1^{(star)}$}\label{i1-star}
	Following theorem establishes the relation between $\mathcal{B}_{n-star}$ and $\mathcal{C}_i$$(i$$=$$1,2,...,n$):
	\begin{theorem}\label{theo4}
		 In any $n$-local star network where $\forall i$$=$$1,...,n$, $\mathcal{S}_i$ distributes arbitrary two-qubit state $\rho_i$ with concurrence $\mathcal{C}_i$, $\mathcal{B}_{n-star}$ and $\mathcal{C}_i$$(i$$=$$1,2,...,n$) satisfy the following relation:
	\begin{equation}\label{ups1}
		\mathcal{B}_{n-star}\leq \sqrt{1+\frac{1}{n}\sum_{i=1}^n\mathcal{C}_i^{2}}
	\end{equation} 
	\end{theorem}

	\textit{Proof:} See Appendix.D\\

	Let Eq.(\ref{ups1}) be denoted as the inequality \textbf{$I_1^{(star)}$}. Analogous to linear $n$-local network, maximum value(with respect to all two-qubit states) of $\mathcal{B}_{n-star}$ is $\sqrt{2}$ and $I_1^{(star)}$ indicates that such amount of violation is only possible if all the sources are maximally entangled.\\
	R.H.S. of $I_1^{(star)}$ clearly points out the following no-go result for star network.\\
\textit{Corollary.2:}\label{cor2} Non $n$-locality cannot be detected via violation of $I_{n-star}$ if none of the sources distribute entangled state in the star network.\\
However, R.H.S. of Eq.(\ref{ups1}) exceeds $1$ even if one of the sources generate entangled state($\mathcal{C}_i$$>$$0$ for some $i$$\in$$\{1,2,...,n\}).$ Hence, violation may be possible even when all the sources are not entangled. \\
To this end, it may be noted that when all the sources generate two-qubit entangled states the bound provided by $I_1^{(star)}$ can be further refined.
	\begin{theorem}\label{theo5} 
	In any $n$-local star network where $\forall i$$=$$1,...,n$, $\mathcal{S}_i$ distributes arbitrary two-qubit entangled state $\rho_i$ with concurrence $\mathcal{C}_i$, upper bound on $\mathcal{B}_{n-star}$ is given by:
	\begin{equation}\label{upsr1}
		\mathcal{B}_{n-star}\leq \sqrt{1+\Pi_{i=1}^n\mathcal{C}_i^{\frac{2}{n}}}
	\end{equation} 
	\end{theorem}
	\textit{Proof:} See Appendix.E. \\
	Comparing R.H.S. of Eqs.(\ref{ups1},\ref{upsr1}) one can easily see that Eq.(\ref{upsr1}) provides better upper bound of $B_{n-star}.$ $I_1^{(star)}$ and its refined version(Eq.(\ref{upsr1})) coincide only when all the sources have same concurrence. It must be noted here that bound provided by Eq.(\ref{upsr1}) does not hold even if one of the sources does not distribute entangled state. Consequently, this bound cannot be used to analyze violation of $I_{n-star}$ when one of the sources is separable(see subsec.\ref{starvslin} and Appendix.E  for more details).\\
	 For instance, consider a $5$-local star network where $\mathcal{S}_1,\mathcal{S}_2,\mathcal{S}_3,\mathcal{S}_4,\mathcal{S}_5$ each generates a maximally entangled two-qubit state and $\mathcal{S}_5$ distribute a separable two-qubit  Werner state:
	\begin{equation}\label{werner}
		\rho_5^{werner}=\mathfrak{v}_5 |\psi^{-}\rangle\langle \psi^{-}|+\frac{1-\mathfrak{v}_5}{4}\mathbb{I}_{2\times2},
	\end{equation}
	with $\mathfrak{v}_5$$=$$\frac{1}{4}$ denoting visibility parameter of state $\rho_i.$ \\
	In this case:
	\begin{eqnarray}
		B_{n-star}&=&\sqrt{2(\frac{1}{4})^{\frac{2}{5}}}\nonumber\\
	&=&2^{\frac{1}{10}}>1
	\end{eqnarray}
	 But $\rho_5$ being separable, $\mathcal{C}_5$$=$$0.$  R.H.S of Eq.(\ref{upsr1}) thus turns out to be $1.$ So Eq.(\ref{upsr1}) is violated here. Upper bound provided by Eq.(\ref{upsr1}) thus only holds when all the states used in the star network are entangled. \\
	 In the above example we have considered star network with $5$ sources. When one state is separable, violation of $n$-local inequality can also be obtained in the simplest star network, i.e., for $n$$=$$3.$  \\
	 To give a particular numerical instance, let us consider that $2$ of the sources distribute maximally entangled state whereas the remaining one distribute the following separable Bell diagonal state\cite{nie}:
\begin{eqnarray}\label{svm6}
	\rho_{BD}&=&	\left[ {\begin{array}{cccc}
		\frac{1}{4}	&0&0&\frac{1}{16}\\
		0&\frac{1}{4}	&\frac{3}{16}&0\\
		0&\frac{3}{16}&\frac{1}{4}&0\\
		\frac{1}{16}&0&0&\frac{1}{4}
	\end{array} } \right]\nonumber\\
\end{eqnarray}
	 Correlation tensor of the state(Eq.(\ref{svm6})) is given $\textbf{diag}(\frac{1}{2},\frac{1}{4},0).$ \\
	 So $B_{3-star}$ is given by:
	 \begin{equation}
	 	B_{3-star}=\sqrt{(\frac{1}{2})^{\frac{2}{3}}+(\frac{1}{4})^{\frac{2}{3}}}=1.01332	
	 	 \end{equation}
	 	 From above examples it is obvious that non $n$-locality can be obtained in star network involving few separable sources. However, quite obviously the same is impossible if all the sources are separable. It now becomes important to remark on the count of separable sources in $n$-local star network. Though an affirmative response is beyond the scope of present discussion, we add some related insight in form of the following no-go result.
	 	 \begin{theorem}\label{sepcount}
	 	 \textit{In a $n$-local star network having $m$ number of separable sources, non $n$-locality cannot be detected if $m$$\geq $$\left\lceil\frac{n}{2}\right\rceil.$}
	 	 	 	 \end{theorem}
	 	\textit{Proof:} See Appendix.F.\\
	 	By above theorem, for $n$$=$$3,4$ non $n$-locality cannot be detected if $2$ separable sources are used in corresponding star network. Also example discussed prior to Theorem.\ref{sepcount} ensures that non $n$-locality($n$$=$$3,4$) will be obtained if single separable source is used in the network. Hence up to $n$$=$$4$ we get the following result:
	 	 \begin{theorem}\label{sepcount1}
	 		\textit{For $n$$\leq$$4,$ non $n$-locality can be obtained in star topology if and only if the network involves $m$ number of separable sources with $m$$= $$\left\lceil\frac{n}{2}\right\rceil.$}
	 	\end{theorem}
\subsubsection{Tightness of Eq.(\ref{ups1}) and Eq.(\ref{upsr1})}
	 One can easily check that for the classes of states provided by Theorem.\ref{theo2}, the upper bound provided by Theorem.\ref{theo5} is also satisfied.\\
	 So, for any states prescribed by Theorem.\ref{theo2}:
	 \begin{equation}
	 	B_{n-star}=\sqrt{1+\Pi_{i=1}^n\mathcal{C}_i^{\frac{2}{n}}}
	 \end{equation}
	 Under identical source entanglement, R.H.S. of Eq.(\ref{ups1}) and Eq.(\ref{upsr1}) become same. Consequently, $I_1^{(star)}$ is saturated when each of such sources distributes classes of states provided by Theorem.\ref{theo2}.
	\subsubsection{Upper Bound on Maximum Violation of $I_{n-star}$} Inequality $I_1^{(star)}$ gives an upper limit on the maximum(with respect to measurement parameters) possible violation of $I_{n-star}$ in terms of source entanglement. 	$I_1^{(star)}$(Eq.(\ref{ups1})) implies that violation of $I_{n-star}$ can never exceed $\sqrt{1+\frac{1}{n}\sum_{i=1}^n\mathcal{C}_i^{2}}.$ Consequently, for any given set of concurrences $\mathcal{C}_i(i$$=$$1,2,...,n),$ the maximum violation of $I_{n-star}$ in any $n$-local star network always remain bounded by this quantity.\\
	Now let each of the states used in the network be entangled. Then, for any given amount of violation $V^{'}$(say) of $I_{n-star}$ the source entanglement must satisfy:
	\begin{equation}\label{fv2}
		\mathcal{K}\geq((1+V^{'})^2-1)^{\frac{n}{2}}.
	\end{equation}

	Let $V^{'}$ denote a fixed violation amount of $I_{n-star}$ using entangled states as prescribed by Theorem.\ref{theo2}. In that case the refined version of $\mathcal{I}_1^{(star)}$(Eq.(\ref{upsr1})) is applicable. Using Eq.(\ref{upsr1}) we get:
	\begin{equation}\label{ups2}
		\Pi_{i=1}^n\mathcal{C}_i=((1+V^{'})^2-1)^{\frac{n}{2}}.
	\end{equation}
	When all the sources have same entanglement content, Eq.(\ref{ups2})gives the minimum entanglement content($\mathcal{C}$) required for $V^{'}$:
	\begin{equation}\label{ups3}
		\mathcal{C}=\sqrt{(1+V^{'})^2-1}
	\end{equation}
	\subsection{Derivation of $I_2^{(star)}$ }\label{i2-lin}
	We now derive the lower bound of $B_{n-star}.$ The theorem below provides the same for Bell diagonal class of states.\\
	\begin{theorem}\label{theo6}
			 When $\mathcal{S}_i$ distributes arbitrary Bell-diagonal state $\rho_i$ with concurrence $\mathcal{C}_i$$(i$$=$$1,2,...,n$) in any $n$-local star network, $\mathcal{B}_{n-star}$ is lower bounded by a function of $\mathcal{C}_i$$(i$$=$$1,2,...,n$):
		\begin{equation}\label{ups4}
			\mathcal{B}_{n-star}\geq \sqrt{2}\Pi_{i=1}^n\mathcal{C}_i^{\frac{1}{n}}
		\end{equation}
	\end{theorem}

	\textit{Proof:} Similar to the proof of Theorem.\ref{theo3}\\ 
	Eq.(\ref{ups4}) will be referred to as \textbf{$I_2^{(star)}$}. $I_2^{(star)}$ implies that non $n$-locality can be detected in a star network using Bell diagonal states satisfying the relation(see sub-fig.b in Fig.\ref{figstarlower} for$n$$=$$3$):
	\begin{equation}\label{ups5}
		\Pi_{i=1}^n\mathcal{C}_i>\frac{1}{2^{\frac{n}{2}}}.
	\end{equation} 
	Eq.(\ref{ups5}) aids in characterizing the entanglement content of Bell-diagonal states generating non $n$-locality in star network. Clearly, any collection of Bell-diagonal states violating Eq.(\ref{ups5}) cannot violate $I_{n-star}$.\\
	\par One can easily check that tightness of $I_2^{(star)}$ holds for the same states that saturates  $I_2^{(lin)}$. \\
	So far we have dealt with the linear and star shaped $n$-local networks individually. Using our analysis from last two sections, we next proceed to compare the amount of non $n$-locality that can be detected in these two different types of networks.
	\subsubsection{Applying $I_2^{(star)}$ Beyond Bell-Diagonal Class} 
	For showing applicability of $I_2^{(star)}$ outside the Bell-Diagonal we consider the subclass of X states(Eq.(\ref{xst2})). But doing analysis similar to that made for $I_2^{(lin)},$ one can easily check that $I_2^{(star)}$ also holds for this subclass of X states. \\
	Also, the quantity $\mathcal{B}_{n-star}$$-$$ \sqrt{2}\Pi_{i=1}^n\mathcal{C}_i^{\frac{1}{n}}$ is tested for randomly many set of two-qubit states. Based on the numerical observations(see sub-fig.(b) in Fig.\ref{linrand}) we give the following conjecture:\\
\textit{\textbf{Conjecture.2:} For any set of two-qubit states, $I_2^{star}$ is satisfied.}
	\section{Comparing non $n$-locality In Linear and star networks}\label{starnow1}
	Comparison of $B_{n-linear}$ and $B_{n-star}$ indicates the supremacy of star configuration over linear pattern to detect non $n$-locality in a $n$-local network. 
	\subsection{$B_{n-linear}$ vs $B_{n-star}$}\label{starvslin}
\begin{center}
\begin{figure}
\includegraphics[width=3.4in]{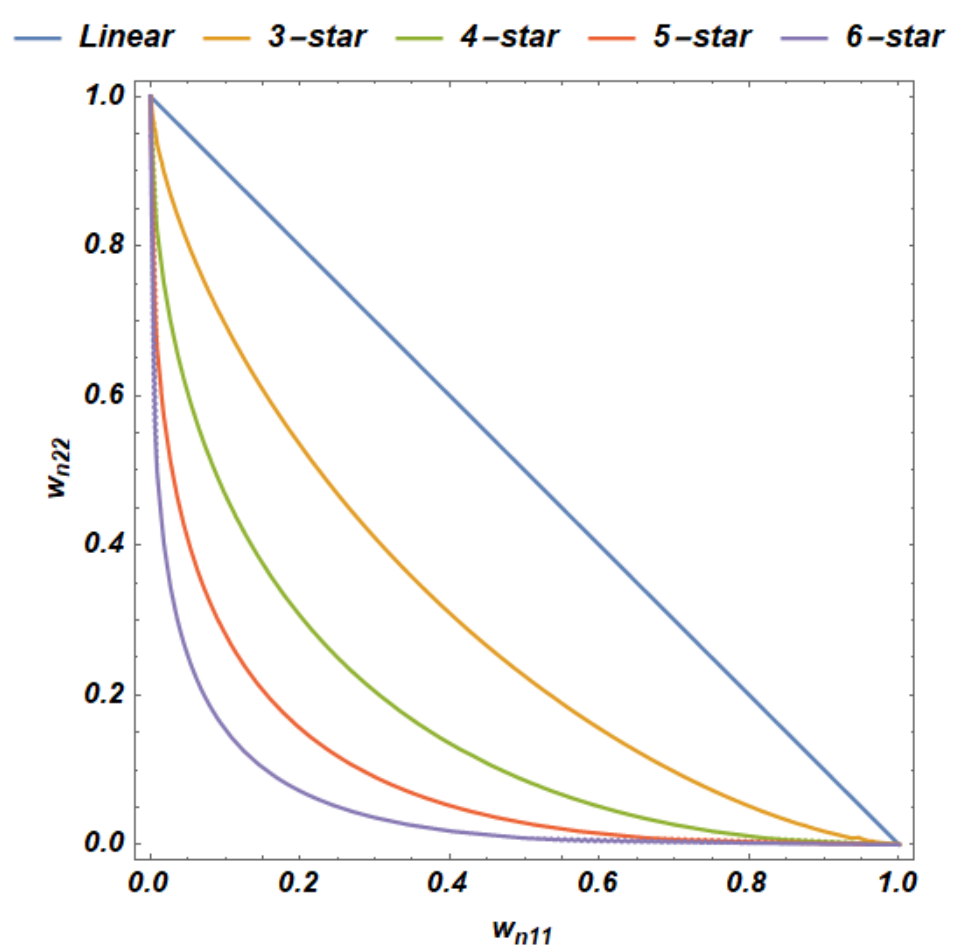} \\
\caption{\emph{Curves are plotted in space of largest two singular values($E_{n1},E_{n2}$) of correlation tensor $W^{(n)}$ corresponding to arbitrary separable two-qubit state($\rho_n$). $\rho_n$ being separable, no region exists outside the outermost curve marked as 'linear'\cite{See}. This curve gives the boundary of region specified by Eq.(\ref{new1}). So, here non $n$-locality cannot be detected in $n$-local linear network. For $n$$=$$3,4,5,6$, space outside curves marked as $n-star$ and bounded by outermost straight line gives region of detectable non $n$-locality in $n$-local star network. Staring from the outermost curve(marked $3-star$), inward bending of the curves indicate that with increasing number of sources, star $n$-local network turns out to be more efficient for detecting non $n$-locality. }}
\label{wercomps1}
\end{figure}
\end{center}

	Let us consider the case when $n$$-$$1$ states $\rho_1,\rho_2,...,\rho_{n-1}$ are maximally entangled and only a single source is separable. If $\rho_i$ is maximally entangled state then $|E_{i1}|$$=$$|E_{i2}|$$=$$|E_{i3}|$$=$$1.$ Then:
	\begin{eqnarray}\label{new1}
		B_{n-linear}&=&\sqrt{|E_{n1}|+|E_{n2}|}\nonumber\\
			&\leq& 1
	\end{eqnarray}
	Inequality in last line of Eq.(\ref{new1}) follows from the separability criterion for two-qubit states given in \cite{See}. \\
	However, in star network, we get:
	\begin{equation}\label{new2}
		B_{n-star}=\sqrt{|E_{n1}|^{\frac{2}{n}}+|E_{n2}|^{\frac{2}{n}}}
	\end{equation}
	Comparison Eqs.(\ref{new1},\ref{new2}) points out that $\forall\, n$$\geq$$3,$ $B_{n-star}$ exceeds $B_{n-linear}.$ Hence, there exist separable states, which when used with maximally entangled states, can show non $n$-locality in star network(see Fig.\ref{wercomps1}). Such an example is already provided in subsec.\ref{i1-star}.\\
	It is clear that whenever $m$$<$$n$ number of sources are separable and $n$$-$$m$ are entangled, star network may exhibit non $n$-locality in contrast to linear $n$-local network for suitable choice of measurement settings. 
	We intend to compare detectable amount of non $n$-locality across linear and star shaped $n$-local networks in terms of entanglement content of the independent sources.
	\subsection{Measure of Non $n$-locality}
	While concurrence will be used for entanglement content, we use the following quantity as a measure of non $n$-locality in a network with topology $\mathcal{T}:$
	\begin{equation}\label{meas1}
		M_{\mathcal{T}}=\frac{V_{\mathcal{T}}}{V_{\mathcal{T}}^{(max)}}
	\end{equation}
	In Eq.(\ref{meas1}), $V_{\mathcal{T}}$ represent the maximum violation amount of a $n$-local inequality with respect to measurement variables. $V_{\mathcal{T}}^{(max)}$ denote the maximum possible value of $V_{\mathcal{T}}$ with respect to all possible sets of $n$ number of two-qubit states. Clearly, $M_{\mathcal{T}}$$\in$$[0,1].$ For star and linear topologies, $V_{\mathcal{T}}^{(max)}$$=$$0.414.$ Measure of non $n$-locality(Eq.(\ref{meas1})) in linear and star network explicitly takes the form:
	\begin{eqnarray}\label{meas2}
		M_{n-linear}&=&\textmd{Max}[0,\frac{V_{linear}}{0.414}]\\
		M_{n-star}&=&\textmd{Max}[0,\frac{V_{star}}{0.414}]
	\end{eqnarray}
	Using above measures, let us consider the following term:
	\begin{equation}\label{diff}
		\mathcal{D}_n=\mathcal{M}_{n-star}-\mathcal{M}_{n-linear}
	\end{equation}
	$\mathcal{D}_n$$>$$0$ thus acts as an indicator of larger efficiency of star network to exploit non $n$-locality compared to linear topology. Under identical source entanglement content in both the networks, any positive amount of $\mathcal{D}_n$ quantifies the advantage of star network over the linear one.
	\subsection{Comparison Between Linear and Star Network}\label{compa}
\begin{center}
	\begin{figure}
		\includegraphics[width=3.4in]{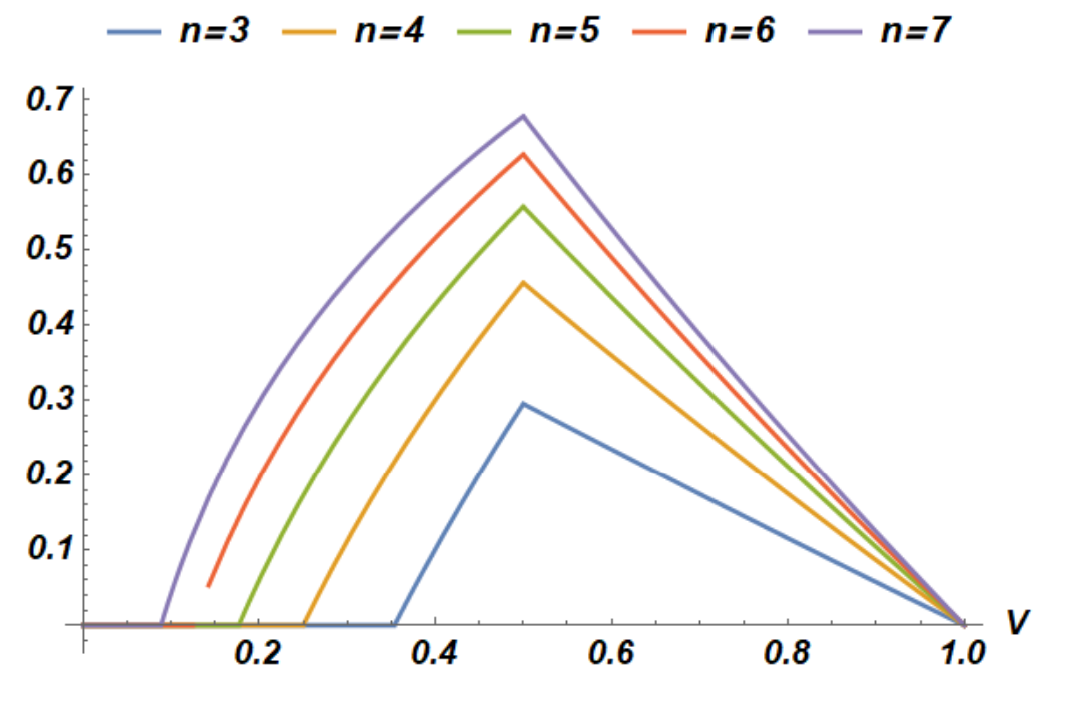} \\
		\caption{\emph{$\mathcal{D}_{n}$(for $n$$=$$3,4,5,6,7)$  is plotted against the product($V$) of visibility($v_i$) of the Werner states($\rho_i^{(werner)}$). Clearly, in context of generating detectable non $n$-local correlations, star network provides advantage($\mathcal{D}_n$$>$$0$) over linear network for some range of $V.$ }}
		\label{wercomps}
	\end{figure}
\end{center}

	Let us start with the example of Werner states, i.e., $\forall n$$=$$1,2,...,n,$ let $\mathcal{S}_i$ distribute a two-qubit Werner state $\rho_i$(Eq.(\ref{werner}) with $v_5$ replaced by $v_i$). When these sources are used in linear and star network, we get:
	\begin{eqnarray}\label{new3}
\mathcal{D}_n&=& 0\,\small{\textmd{if}}\,V\leq\frac{1}{2^{\frac{n}{2}}} \nonumber\\
&=&3.41597 V^{\frac{1}{n}} \,\small{\textmd{if}}\,\frac{1}{2^{\frac{n}{2}}}<V\leq\frac{1}{2}\nonumber\\
&=&3.41597(V^{\frac{1}{n}}-V^{\frac{1}{2}})\, \small{\textmd{else}}\\
&&\small{\textmd{where}}\, V=\Pi_{i=1}^nv_i
\end{eqnarray}
Clearly, $\mathcal{D}_n$ turns out to be positive over some range of $V$(see Fig.\ref{wercomps}). This in turn points out efficiency of star network for generating non $n$-locality.
For a specific example, let us consider $4$ Werner states $\rho_i^{(werner)}(i$$=$$1,2,3,4)$ with visibility parameters as follows:$v_1$$=$$0.28$ and $v_2$$=$$v_3$$=$$v_4$$=$$0.97.$ Eq.(\ref{new3}) gives $\mathcal{D}_4$$=$$0.0133.$
Till now, we have compared between the two network topologies in the most general case where the sources generate any arbitrary two-qubit state.Next, we narrow down our discussion to the case where only entangled two-qubit states are used in the networks. For that, findings related to $I_1^{(linear)}$ and finer version of $I_1^{(star)}$ will be used for obvious reasons.  
	
	\subsubsection{Comparing Non $n$-locality  for Fixed Source Entanglement}
\begin{center}
		\begin{figure}
			\includegraphics[width=3.4in]{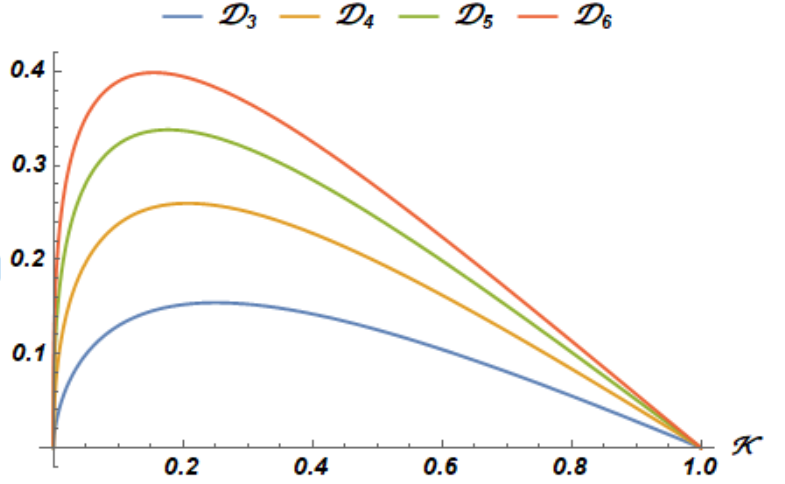} \\
			\caption{\emph{$\mathcal{D}_n$ is plotted against $\mathcal{K}$ for $n$$=$$3,4,5,6$ number of sources. Clearly, the difference between the amount of non $n$-locality increases with increasing number of sources. }}
			\label{figcomp}
		\end{figure}
	\end{center}

	Let $\mathcal{S}_1,...,\mathcal{S}_n$ be $n$ independent sources. Let $\mathcal{S}_i$ distribute any two-qubit entangled state $\rho_i$  with concurrence $\mathcal{C}_i$ as specified in Theorem.\ref{theo2}. As said before, $\mathcal{K}$ characterizes entanglement of the sources used in the network.\\
	 Let violation of $I_{n-linear}$ is observed when these sources are arranged in a linear pattern. Using Eq.(\ref{up2}), amount of non $n$-locality observed in the linear network is given by:
	\begin{equation}\label{meas3}
		M_{n-linear}=\frac{\sqrt{1+\mathcal{K}}-1}{0.414}
	\end{equation}
	Now, let the same sources be arranged in a star-shaped arrangement. By  Eq.(\ref{ups2}), amount of non $n$-locality observed in the star network is given by:
	\begin{equation}\label{meas4}
		M_{n-star}=\frac{\sqrt{1+\mathcal{K}^{\frac{2}{n}}}-1}{0.414}
	\end{equation} 
	 Comparing Eqs.(\ref{meas3},\ref{meas4}), it is clear that for any given amount of entanglement in the sources, $M_{n-star}$$\geq$$M_{n-linear}$. Hence, more non $n$-locality can be detected($\mathcal{D}_n$$\geq$$0$) in star network compared to that in linear network(see Fig.\ref{figcomp}). For $\mathcal{K}$$=$$0,1$ same amount of non $n$-locality(hence $\mathcal{D}$$=$$0$) can be detected in both star and linear networks. 
	\subsubsection{Comparing Source Entanglement for fixed amount of Non $n$-locality}
\begin{center}
		\begin{figure}
			\includegraphics[width=3.4in]{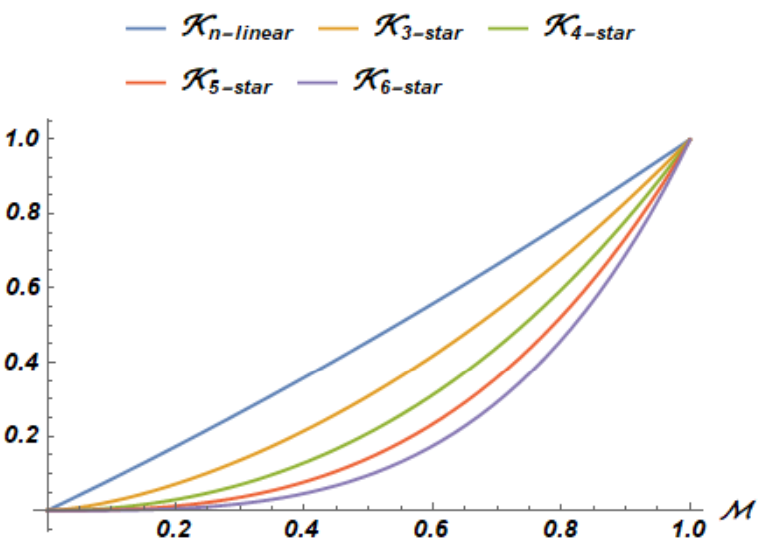} \\
			\caption{\emph{$\mathcal{K}_{n-linear},$ $\mathcal{K}_{n-star}$(for $n$$=$$3,4,5,6)$  is plotted against some fixed value $\mathcal{M}$ of measure of non $n$-locality. Clearly, lesser amount of entanglement is required in star network compared to linear network.}}
			\label{figcomp2}
		\end{figure}
	\end{center}

	Conversely, let $\mathcal{M}$ denote a fixed amount of non $n$-locality detected in both linear and star network. Let $\mathcal{K}$$=$$\mathcal{K}_{n-linear}$ and $\mathcal{K}_{n-star}$ parameterizing source entanglement for linear and star networks respectively exhibiting $\mathcal{M}$ amount of non $n$-locality. Then using Eqs.(\ref{meas3},\ref{meas4}), we get:
	\begin{eqnarray}\label{meas5}
		\mathcal{K}_{n-linear}&=&(0.414 \mathcal{M}_{n-linear}+1)^2-1\\
		\mathcal{K}_{n-star}&=&((0.414 \mathcal{M}_{n-star}+1)^2-1)^{\frac{n}{2}}
	\end{eqnarray}
	Eq.(\ref{meas5}) points out that for any fixed $\mathcal{M},$ entanglement requirement of sources in star network is less(and gradually decreases with increase in $n$) compared to that in linear network(see Fig.\ref{figcomp2}).
	\section{Practical Implications}\label{prac}
	Our discussion has so far illustrated the interplay between source entanglement with that of generation of $n$-local inequality violation in a $n$-local network. In this section, we intend to point out few practical implications of our findings. 
	\subsection{Entanglement Detection In Networks}
	Consider a $n$-local linear network with $n$ independent sources. Let each of $\mathcal{S}_1,...,\mathcal{S}_n$ be uncharacterized in the sense that there is no information about the entanglement content of the physical states distributed from them across the network. Let the parties perform measurements as stated before in sec.\ref{pre}. Resulting measurement correlations are collected and then $I_{n-linear}$ is tested. Let violation of $I_{n-linear}$ is observed. By corollary.\ref{cor1}, each of the sources must be entangled. Corollary.\ref{cor1} thus turns out to be effective in detecting entanglement of all sources in $n$-local linear network. Similarly Corollary.\ref{cor2} aids in detecting entanglement of sources in $n$-local star network. 
	\subsection{Robustness To Noise}
	Ideally pure entangled states are supposed to be distributed across a network. However, due to environmental interactions, noisy entangled states get distributed among the nodes. So for practical situations it becomes important to use network scenario that  is more resistant to environmental noise. From this perspective,  discussions in sec.\ref{compa} point out the greater effectiveness of star network compared to linear scenario. To be precise, for any given set of noisy two qubit entangled states satisfying conditions of Theorem.\ref{theo2}, $\mathcal{D}_n$$>$$0.$ Hence, correlations generated in star network will have higher noise resistance compared to linear network.\\
	Let us next consider states that do not satisfy conditions in Theorem.(\ref{theo2}). For example, Werner class of states(Eq.(\ref{werner})) neither belong to class of pure states nor to family of Rank-$2$ Bell-Diagonal states. This class of noisy two-qubit states result due to communication of two-qubit singlet states over global depolarization channel\cite{nie}. 
	Discussions, given in subsec.\ref{compa} points out the existence of range of visibility parameters $v_i$for which non $n$-locality is detectable only in star network but not in linear network(see Fig.\ref{wercomps}). \\
	Let us now consider that only a single source distribute Werner state whereas each of remaining $n-1$ sources distribute a two-qubit maximally entangled state. W.L.O.G., let $\mathcal{S}_1$ distribute Werner state(with visibility $\mathfrak{v}_1$). \\
	Non $n$-locality can be obtained in star network if:
	
		\begin{eqnarray}\label{faltu}
		B_{n-star}&=&\sqrt{2\mathfrak{v}_1^{\frac{2}{n}}}>1\nonumber\\
		\mathfrak{v}_1&>&\frac{1}{2^{\frac{n}{2}}}
	\end{eqnarray}
Eq.(\ref{faltu}) indicates that with increase in number of edge parties($n$), the range of visibility $\mathfrak{v}_1$ of Werner state for which non $n$-locality can be exploited, increases. 
\subsection{Detecting Variable Non-classicality For Fixed Source Entanglement}
In any network scenario, the term $\kappa$$=$$\Pi_{i=1}^n$ characterizes entanglement content involved across the entire network structure. Let us suppose that for some network based information processing task we need to construct a linear network having $\kappa$$=$$\kappa_{lin}.$ For that we can choose to use several sets of $n$ number of two-qubit states $\{\rho_1,\rho_2,...\rho_n\}$ in the network such that the product of their concurrences is $\kappa_{lin}.$ For our purpose, let us suppose that we have chosen the following sets:
\begin{itemize}
	\item  $\mathcal{S}^{'}$$=$$\{\rho_1^{'},\rho_2^{'},...\rho_n^{'}\}$ such that this class of states saturates $\mathcal{I}_1^{(lin)}.$
	\item $\mathcal{S}^{''}$$=$$\{\rho_1^{''},\rho_2^{''},...\rho_n^{''}\}$ such that this class of states violates $\mathcal{I}_{n-linear}$ and saturates $\mathcal{I}_2^{(lin)}.$
\end{itemize}
Let $\mathcal{M}_{n-linear}^{'}$, $\mathcal{M}_{n-linear}^{''}$ be the respective amount of non $n$-locality(using Eq.(\ref{meas2})) obtained when we use $\mathcal{S}^{'},$ $\mathcal{S}^{''}$ in the linear network. By the properties of $\mathcal{I}_1^{(lin)}$ and $\mathcal{I}_2^{(lin)}$ we have:
\begin{eqnarray}
	\mathcal{M}_{n-linear}^{'}-\mathcal{M}_{n-linear}^{''}&=&\frac{\sqrt{1+\kappa}
	-\sqrt{2\kappa}}{0.414}\nonumber\\
	&\geq& 0
\end{eqnarray}
The case of equality is possible only when $\kappa$$=$$1$ but such a situation is again practically infeasible for obvious reasons. So more non $n$-locality is obtained in linear network using $\mathcal{S}^{'}$ compared to $\mathcal{S}^{''}.$ Clearly, same is the case if we consider the $n$-local star network. Thus, for any fixed budget for entanglement cost, it will be more beneficial to use $\mathcal{S}^{'}$ compared to $\mathcal{S}^{''}$ in $n$-local network in general. Let us illustrate with a specific numerical example.\\
Let us consider a $3$-local network. Let us fix $\kappa$$=$$0.576.$ Consider a set of $3$ states($\mathcal{S}^{''}$) given by Eq.(\ref{horodeck}) for $p_1$$=$$p_2$$=$$0.8$ and $p_3$$=$$0.9.$
Consider another set($\mathcal{S}^{'}$) of $3$ states as follows:
\begin{eqnarray}
		\rho_1^{'}=\frac{1}{2} \begin{pmatrix}
		0& 0 & 0 & 0\\
		0& 0.6028 & 0.8&0 \\
		0&  0.8 & 1.3972&0 \\
		0& 0 &  0& 0\\
	\end{pmatrix}\nonumber\\
		\rho_2^{'}= \frac{1}{2} \begin{pmatrix}
		0& 0 & 0 & 0\\
		0& 0.6657 & 0.8&0 \\
		0&  0.8 & 1.3343&0 \\
		0& 0 &  0& 0\\
	\end{pmatrix}\nonumber\\
		\rho_3^{'}=\frac{1}{2}  \begin{pmatrix}
		0& 0 & 0 & 0\\
		0& 0.7211 & 0.8&0 \\
		0&  0.9 & 1.2789&0 \\
		0& 0 &  0& 0\\
	\end{pmatrix}\nonumber\\
\end{eqnarray}
 When we use above $\mathcal{S}^{'}$ and $\mathcal{S}^{''}$ in linear $3$-local network, non trilocality measures are given by:
 \begin{eqnarray}\label{extrap1}
 \mathcal{M}_{3-linear}^{'}&=&0.6169\nonumber\\
 \mathcal{M}_{3-linear}^{''}&=&0.1771
 \end{eqnarray}
 On using $\mathcal{S}^{'}$ and $\mathcal{S}^{''}$ in $3$-local star network, non trilocality measures are given by:
  \begin{eqnarray}\label{extrap2}
 	\mathcal{M}_{3-star}^{'}&=&0.7268\nonumber\\
 	\mathcal{M}_{3-star}^{''}&=&0.4267
 \end{eqnarray}
Eqs.(\ref{extrap1},\ref{extrap2}) clearly point out that $\mathcal{S}^{'}$ will exhibit more non tri-locality compared to $\mathcal{S}^{'}$ in both linear and star topologies. \\
For any fixed resource of entanglement across a network, one can thus segregate class of states which differ in exploiting network non $n$-locality irrespective of network topology. 
\section{Conclusion}\label{conc}
	From what has been discussed so far, one can safely conclude that non $n$-locality can be exploited in star network even when all the sources are not entangled. However the same is not true if the sources are arranged in a linear fashion. \\
	Broadly speaking, present study has focused on illustrating inter-relationship between entanglement of sources and non $n$-locality detection in $n$-local linear and star network scenarios. For detecting non $n$-locality violation of existing $n$-local inequalities($I_{n-linear}$ for linear and $I_{n-star}$ for star topology)
	have been used. In terms of concurrence of the two-qubit states, both upper and lower bounds for each of $I_{n-linear}$ and $I_{n-star}$ have been derived. Utilizing those bounds, several interesting findings have emerged. It is observed that entanglement of each of $n$ independent sources is mandatory for detecting non $n$-locality. Also we have characterized sources in terms of minimum entanglement requirement for any fixed amount of violation of both $I_{n-linear}$ and $I_{n-star}.$ Product of concurrences of the sources greater than $\frac{1}{2}$ turns out to be sufficient for detecting non $n$-locality. In course of analysis, few instances of two-qubit states showing extremal violations of $I_{n-star}$ and $I_{n-linear}$ have been provided. \\
	\par Our study points out the effectiveness of star $n$-local network over linear network for detecting non $n$-local correlations. Using a measure of non $n$-locality, it is observed that for any given amount of source entanglement, more non $n$-locality is generated in star network. Conversely, given any fixed amount of non $n$-locality, lesser entanglement of the sources is required in case the sources are used in a star network. \\
	Entire analysis made is based on qubit systems. It will be interesting to extend the same for higher dimensional quantum systems. Due to increasing complexity in network based information processing tasks, establishing relationship between source entanglement and network non $n$-locality beyond star topology\cite{birev} is important for future work. Besides, exploiting relationship between non $n$-locality and other forms of correlation features of the sources deserves attention. As discussed, in star network entanglement of all the sources is not a mandate for generation of non $n$-local correlations. In this context we have only been able to point out presence of how many number of separable two-qubit states renders the $n$-local star network ineffective for generating non $n$-locality(Theorem.\ref{sepcount}). But it will be interesting to seek response to the converse problem:\textit{at most how many number of separable sources can be included in star topology generating non $n$-locality($n$$\geq$$5$)?}
	
	
	\section*{Appendix.A}\label{appna}
	\textit{Proof of Theorem.\ref{theo1}:} $\forall i,$ let each of $\mathcal{S}_i$ distribute an arbitrary two qubit state $\rho_i$(Eq.(\ref{st41})). Let concurrence of $\rho_i$ be $\mathcal{C}_i.$\\
	Let measurement settings of $A_1$ be: $\vec{a}.\vec{\sigma}$ and  $\vec{a}^{'}.\vec{\sigma}$ with $\vec{a}$$=$$(a_1a_2,a_3)$,$\vec{a}^{'}$$=$$(a_1^{'},a_2^{'},a_3^{'})$$\in$$\mathfrak{R}^3.$\\
	Similarly let measurement settings of $A_{n+1}$ be: $\vec{b}.\vec{\sigma}$ and  $\vec{b}^{'}.\vec{\sigma}$
	with $\vec{b}$$=$$(b_1b_2,b_3)$,$\vec{b}^{'}$$=$$(b_1^{'},b_2^{'},b_3^{'})$$\in$$\mathfrak{R}^3.$\\
	With these measurement settings of $A_1$ and $A_{n+1}$ and with each of $A_2,...,A_n$ performing projection in Bell basis, the correlator term $I_n$ in linear $n$-local inequality(Eq.(\ref{ineqb})) is given by:
	\begin{eqnarray}\label{extr1}
		I_n&=&\frac{1}{4}(\textmd{Tr}[(\vec{a}+\vec{a}^{'}).\vec{\sigma}\times\sigma_3\times\sigma_3\times....\nonumber\\&&
		\times\sigma_3(2n-2\, times)\times (\vec{c}+\vec{c}^{'}).\vec{\sigma}.\times_{j=1}^n\rho_j])
	\end{eqnarray}

	Any two qubit state $\rho_i$ with concurrence $\mathcal{C}_i$ can be written as convex combination of pure two-qubit states each having concurrence $\mathcal{C}_i$\cite{Woot}:
	\begin{equation}\label{app1}
			\rho_i=\sum_{j=1}^t p_{ij}|\phi_{i,j}\rangle\langle \phi_{i,j}|,\forall i=1,2,...,n
	\end{equation}
	Up to local unitaries each of $|\phi_{i,j}\rangle$ can be written as(Schmidt decomposition\cite{nie}):
	\begin{eqnarray}\label{app15}
		|\phi_{i,j}\rangle&=&\nu_{i0} |00\rangle+\nu_{i1}|11\rangle\nonumber\\
		\textmd{with}\, \nu_{ik}&=&\frac{\sqrt{1+\mathcal{C}_i}+(-1)^k \sqrt{1-\mathcal{C}_i}}{2}\nonumber
	\end{eqnarray}
	As entanglement content remains the same under local unitary operations, so one can consider Eq.(\ref{app15}) for expressing $\rho_i\forall i.$\\
$\forall i,j$ let $W^{(ij)}$ denote the correlation tensor of $|\phi_{i,j}\rangle\langle \phi_{i,j}|. $
	
	From Eq.(\ref{extr1},\ref{app1}), we get:
	\begin{eqnarray}\label{extr2}
		I_n&=&\frac{1}{4}((\sum_{k=1}^3(a_k+a_k^{'})\sum_{m_1=1}^{4}W^{(1m_1)}_{3k})
		(\sum_{m_2=1}^{4}W^{(2m_2)}_{33}).....\nonumber\\	&&(\sum_{m_n=1}^{4}W^{(m\,m_n)}_{3t}\sum_{k=1}^3(b_t+b_t^{'})))
	\end{eqnarray}
	Next, let us focus on the terms of the correlation tensor $W^{(ij)}\forall i,j.$ $W^{(ij)}$ is given by:
	\begin{eqnarray}\label{app2}
		W^{(ij)}&=&\textmd{diag}(\delta_{1,i,j}\mathcal{C}_i^{\xi_{1ij}},\delta_{2,i,j}\mathcal{C}_i^{\xi_{2ij}},\nonumber\\
		&&\delta_{3,i,j}\mathcal{C}_i^{\xi_{3ij}})
	\end{eqnarray}
	where $\delta_{i_1,i_2,i_3}$$\in$$\{1,-1\}$ for $i_1$$=$$1,2,3,$ $i_2$$=$$1,2,...,n,$   $i_3$$=$$1,2,...,t$ $\xi_{1ij},\xi_{2ij},\xi_{3ij}$$\in$$\{1,0\}$ such that $\xi_{1ij}$$+$$\xi_{2ij}$$+$$\xi_{3ij}$$=$$2.$
	
	Using Eqs.(\ref{app1},\ref{app2}), up to local unitary equivalence, correlation tensor($W^{(i)}$) of $\rho_i$ given by:
	\begin{eqnarray}\label{app3}
		W^{(i)}&=&\sum_{j=1}^t p_{ij} W^{(ij)}\nonumber\\
		&=&\textmd{diag}(\sum_{j=1}^t p_{ij}\delta_{1,i,j}\mathcal{C}_i^{\xi_{1ij}},\sum_{j=1}^t p_{ij}\delta_{2,i,j}\mathcal{C}_i^{\xi_{2ij}},\nonumber\\
		&&\sum_{j=1}^t p_{ij}\delta_{3,i,j}\mathcal{C}_i^{\xi_{3ij}})\nonumber\\
		&&
	\end{eqnarray}
	For any $i,$ $E_{i1}$$\geq$$E_{i2}$$\geq$$E_{i3}$ denote the eigen values of $\sqrt{(W^{(i)})^TW^{(i)}}.$
	For any $i,$ all possibilities of $E_{i1},$ $E_{i2},$ $E_{i3}$ are as follows:
	\begin{eqnarray}\label{app4}
		(E_{i1},E_{i2},E_{i3})&=&(\sum_{j=1}^t p_{ij}\delta_{1,i,j}\mathcal{C}_i^{\xi_{1ij}},\sum_{j=1}^t p_{ij}\delta_{2,i,j}\mathcal{C}_i^{\xi_{2ij}},\nonumber\\
		&&\sum_{j=1}^t p_{ij}\delta_{3,i,j}\mathcal{C}_i^{\xi_{3ij}})\nonumber\\
		(E_{i1},E_{i2},E_{i3})&=&(\sum_{j=1}^t p_{ij}\delta_{2,i,j}\mathcal{C}_i^{\xi_{2ij}},\sum_{j=1}^t p_{ij}\delta_{1,i,j}\mathcal{C}_i^{\xi_{1ij}},
		\nonumber\\
		&&\sum_{j=1}^t p_{ij}\delta_{3,i,j}\mathcal{C}_i^{\xi_{3ij}})\nonumber\\
		(E_{i1},E_{i2},E_{i3})&=&(\sum_{j=1}^t p_{ij}\delta_{2,i,j}\mathcal{C}_i^{\xi_{2ij}},\sum_{j=1}^t p_{ij}\delta_{3,i,j}\mathcal{C}_i^{\xi_{3ij}},\nonumber\\
		&&\sum_{j=1}^t p_{ij}\delta_{1,i,j}\mathcal{C}_i^{\xi_{1ij}})\nonumber\\
		(E_{i1},E_{i2},E_{i3})&=&(\sum_{j=1}^t p_{ij}\delta_{3,i,j}\mathcal{C}_i^{\xi_{3ij}},\sum_{j=1}^t p_{ij}\delta_{2,i,j}\mathcal{C}_i^{\xi_{2ij}},\nonumber\\
		&&\sum_{j=1}^t p_{ij}\delta_{1,i,j}\mathcal{C}_i^{\xi_{1ij}})\nonumber\\
		(E_{i1},E_{i2},E_{i3})&=&(\sum_{j=1}^t p_{ij}\delta_{1,i,j}\mathcal{C}_i^{\xi_{1ij}},\sum_{j=1}^t p_{ij}\delta_{3,i,j}\mathcal{C}_i^{\xi_{3ij}},\nonumber\\
		&&\sum_{j=1}^t p_{ij}\delta_{2,i,j}\mathcal{C}_i^{\xi_{2ij}})\nonumber\\
		(E_{i1},E_{i2},E_{i3})&=&(\sum_{j=1}^t p_{ij}\delta_{3,i,j}\mathcal{C}_i^{\xi_{3ij}},\sum_{j=1}^t p_{ij}\delta_{1,i,j}\mathcal{C}_i^{\xi_{1ij}},\nonumber\\
		&&\sum_{j=1}^t p_{ij}\delta_{2,i,j}\mathcal{C}_i^{\xi_{2ij}})	\nonumber\\
		&&
	\end{eqnarray}
	Next let us consider two cases:
	\begin{enumerate}
		\item All states are entangled: $\mathcal{C}_i$$\neq$$0.$
		\item Let at least one of the states be separable. W.L.O.G. let $\mathcal{C}_1$$=$$0.$
	\end{enumerate}
	We deal with these two cases separately.\\
	\textbf{Case.1:} $\forall i$,$\mathcal{C}_i$$\neq$$0:$
	As $\sum_{j=1}^t p_{ij}\delta_{k,i,j}$ denote linear combination of probability terms for $k$$=$$1,2,3$, hence:
	\begin{eqnarray}\label{app5}
		|\sum_{j=1}^t p_{ij}\delta_{k,i,j}\mathcal{C}_i^{\xi_{sij}}|&\leq& \mathcal{C}_i,\,\textmd{if}\, \xi_{sij}=1\forall k,i,\, \textmd{\small{and for some }}s\nonumber\\
		&\leq& 1\,\textmd{else}
	\end{eqnarray}
	Using Eqs.(\ref{app4},\ref{app5}), we get:
	\begin{eqnarray}\label{extr3}
		E_{i1}&\leq&1\nonumber\\
		E_{i2},E_{i3}&\leq& \mathcal{C}_i,\forall i=1,2,...,n
	\end{eqnarray}
	Using Eqs.(\ref{app2},\ref{app5},\ref{extr3}), it is clear that $\forall i,j$ upper bounds of each element of $W^{(ij)}$ is independent of $p_{ij}$ terms. Consequently, the maximization strategy used in \cite{Gisi} is applicable here and the measurement settings maximizing Eq.(\ref{ineqb}) here are same as that used in \cite{Gisi} . Following that we get the result:
	\begin{eqnarray}\label{app5iv}
		B_{n-linear}&=&\sqrt{\Pi_{i=1}^n E_{i1}+\Pi_{i=1}^n E_{i2}}\nonumber\\
		&\leq&\sqrt{1+\Pi_{i=1}^n\mathcal{C}_i}
	\end{eqnarray}
	This proves Theorem.\ref{theo1} in case.1.\\
	\textbf{Case.2:} $\mathcal{C}_1$$=$$0:$ 
	Then Eq.(\ref{app5}) does not hold for $i$$=$$1.$ Eq.(\ref{app5}) gets modified as: 
	\begin{eqnarray}\label{app5ii}
		|\sum_{j=1}^t p_{ij}\delta_{k,i,j}\mathcal{C}_i^{\xi_{sij}}|&\leq& \mathcal{C}_i,\,\textmd{if}\, \xi_{sij}=1\forall k,s,\, \textmd{\small{for some }}s\,\textmd{\small{and} }i\neq 1\nonumber\\
		|\sum_{j=1}^t p_{ij}\delta_{k,i,j}\mathcal{C}_i^{\xi_{sij}}|&\leq& 1,\,\textmd{if}\, \xi_{sij}\neq 1,\,\,\forall k,s,\,\textmd{\small{and} }i\neq 1\nonumber\\
		|\sum_{j=1}^t p_{1j}\delta_{k,i,j}|&\leq& 1,\,\,\forall k\textmd{ and }i=1
	\end{eqnarray}
	Let for $i$$=$$1:$
	\begin{eqnarray}
			|\sum_{j=1}^t p_{ij}\delta_{k,i,j}|&=&P_k,\,\,k=1,2,3
	\end{eqnarray} 
For $i$$=$$1,$ let
\begin{eqnarray}
	|\sum_{j=1}^t p_{ij}\delta_{k,i,j}|&=&P_k\leq 1,\,\,k=1,2,3
\end{eqnarray}	
W.L.O.G., let $E_{1k}$$=$$P_k.$\\
 Using Eq.(\ref{app5ii}), we thus get:
	\begin{eqnarray}\label{extr3ii}
		E_{i1}&\leq&1,\,\,\forall i=2,...,n\,\nonumber\\
		E_{i2},E_{i3}&\leq& \mathcal{C}_i,\,\forall i=2,...,n\, \textmd{and}\nonumber\\
		P_1,P_2,P_3&\leq&1.
	\end{eqnarray}
Also, for any two-qubit separable state\cite{See}, we have:
\begin{equation}\label{separable}
	P_1+P_2\leq 1
\end{equation}
Using Eqs.(\ref{app5ii},\ref{extr3ii},\ref{separable}) we get following bound over $B_{n-linear}:$
	\begin{eqnarray}\label{separable2}
		B_{n-linear}&=&\sqrt{\Pi_{i=1}^n E_{i1}+\Pi_{i=1}^n E_{i2}}\nonumber\\
		&\leq&\sqrt{P_1.\Pi_{i=2}^n E_{i1}+P_2.\Pi_{i=2}^nE_{i2}}\nonumber\\
		&\leq&\sqrt{P_1+P_2}\,\textmd{as } E_{ij}\leq 1\\
		&\leq& 1,\,\textmd{By Eq.(\ref{separable})}
	\end{eqnarray}
		This shows that when one of $\rho_1,\rho_2,..\rho_n$ is separable then $B_{n-linear}$ does not exceed $1.$ This is true whenever any number of $\rho_1,\rho_2,..\rho_n$ are separable. \\
		In all these situations(at least one of $\rho_1,\rho_2,..\rho_n$ separable), the form of the upper bound claimed in the theorem holds trivially as:
		\begin{eqnarray}\label{separable3}
		B_{n-linear}&\leq&1\nonumber\\
		&=&\sqrt{1+0}\nonumber\\
		&=&\sqrt{1+\Pi_{i=1}^n\mathcal{C}_i}\,\textmd{as }\mathcal{C}_1=0\,(\textmd{\small{W.L.O.G. let }}\rho_1\,\textmd{\small{be separable}})\nonumber\\
		&&
	\end{eqnarray}
This completes the proof of the Theorem. 
	\section*{Appendix.B}
	\textit{Proof of Theorem.\ref{theo2}:}\\
	\textbf{Case.1:} Let $\rho_1,\rho_2,...,\rho_n$ be all pure two qubit states with concurrence $\mathcal{C}_1,\mathcal{C}_2,...,\mathcal{C}_n$. As already discussed above in Appendix.A, correlation tensor of any two-qubit pure state is given by:
	\begin{equation}\label{app7}
		W^{(i)}=\textmd{diag}(\delta_{1,i}\mathcal{C}_i,\delta_{2,i}\mathcal{C}_i,\delta_{3,i})\,
		\delta_{j,i}\in\{+1,-1\}
	\end{equation}
	
	Using Eqs.(\ref{boundlin},\ref{app7}), $B_{n-linear}$ turns out to be:
	\begin{equation}\label{app7i}
		B_{n-linear}=\sqrt{1+\Pi_{i=1}^n\mathcal{C}_i}
	\end{equation}
	Hence, $I_1^{(lin)}$ is saturated if all states used the network are pure two qubit states.\\
	\textbf{Case.2:} Let $\rho_1,\rho_2,...,\rho_n$ be all rank-2 Bell-Diagonal states with concurrence $\mathcal{C}_1,\mathcal{C}_2,...,\mathcal{C}_n$. Up to local unitaries any such density matrix can be written as \cite{Woot}:
	\begin{eqnarray}\label{controlu}
		\rho_i= \frac{1}{2}\begin{pmatrix}
			0& 0 & 0 & 0\\
			0& 1-S_i & \mathcal{C}_i&0 \\
			0&  \mathcal{C}_i & 1+S_i&0 \\
			0& 0 &  0& 0\\
		\end{pmatrix}\nonumber\\
	\textmd{with } |S_i|&<& \sqrt{1-\mathcal{C}_i^2}
	\end{eqnarray}
	Correlation tensor of $\rho_i$(Eq.(\ref{controlu})) is given by:
	\begin{equation}\label{app8}
		W^{(i)}=\textmd{diag}(\mathcal{C}_i,\mathcal{C}_i,-1)
	\end{equation}
	Hence,in this case also $B_{n-linear}$ is given by Eq.(\ref{app7}). \\
	\textbf{Case.3:}Let some of $\rho_1,\rho_2,...,\rho_n$ be pure two qubit states whereas remaining are rank-2 Bell-diagonal states. Then the result follows directly by combining above two cases.\\
	This proves Theorem.\ref{theo2}.
	\section*{Appendix.C}\label{appnc}
	\textit{Proof of Theorem.\ref{theo3}:}
	Let $\rho_1,...,\rho_n$ be two-qubit Bell-Diagonal states. Let $\forall i,$ singular values (arranged in decreasing order) of the correlation tensor of $\rho_i$ be given by $(E_{i1},E_{i2},E_{i3}).$ As already discussed in sec.\ref{pre}, for any two qubit state $B_{n-linear}$ is given by Eq.(\ref{boundlin}):
	\begin{eqnarray}\label{appc1}
		B_{n-linear}&=&\sqrt{\Pi_{i=1}^nE_{i1}+\Pi_{i=1}^n E_{i2}}\nonumber\\
		&\geq&\sqrt{2\sqrt{(\Pi_{i=1}^nE_{i1})(\Pi_{i=1}^nE_{i2})}}\,A.M.\geq G.M.\nonumber\\
		&=&\sqrt{2\sqrt{\Pi_{i=1}^nE_{i1}E_{i2}}}\nonumber\\
		&\geq& \sqrt{2\sqrt{\Pi_{i=1}^n\Pi_{j=1}^3E_{ij}}}\,\small{as}\,E_{ij}\leq 1\nonumber\\
		&=&\sqrt{2\sqrt{ \Pi_{i=1}^n Y_i}}\,\small{where}\, Y_i=\Pi_{j=1}^3E_{ij}
	\end{eqnarray}
	Now, for any Bell-Diagonal state $\rho_i$ with concurrence $\mathcal{C}_i$\cite{bfei}:
	\begin{equation}\label{appc2}
		Y_i\geq \mathcal{C}_i^2\,\,\forall i
	\end{equation}
	Using Eq.(\ref{appc2}) in Eq.(\ref{appc1}), we get:
	\begin{equation}
		B_{n-linear}\geq \sqrt{2 \Pi_{i=1}^n \mathcal{C}_i}
	\end{equation}
	This proves Theorem.\ref{theo3}.
			\section*{Appendix.D}\label{appnd}
		\textit{Proof of Theorem.\ref{theo4}:} $\forall i,$ let each of $\mathcal{S}_i$ distribute an arbitrary two qubit state $\rho_i$(Eq.(\ref{st41})) with concurrence  $\rho_i$ be $\mathcal{C}_i.$\\
		By Eq.(\ref{boundstar}), we get:
		\begin{eqnarray}\label{amproof}
			B_{n-star}&=&\sqrt{(\Pi_{j=1}^n E_{j1})^{\frac{2}{n}}+(\Pi_{j=1}^nE_{j2})^{\frac{2}{n}}}\nonumber\\
			&\leq&\sqrt{\frac{\sum_{j=1}^n E_{j1}^{2}+\sum_{j=1}^nE_{j2}^{2}}{n}},\,(By\,A.M.\geq G.M.)\nonumber\\
			&=&\sqrt{\frac{\sum_{j=1}^n(E_{j1}^{2}+E_{j2}^2)}{n}}\nonumber\\
			&\leq & \sqrt{\frac{\sum_{j=1}^n(1+\mathcal{C}_j^2)}{n}}\nonumber\\
			&=&\sqrt{1+\frac{\sum_{j=1}^n\mathcal{C}_j^2}{n}}
		\end{eqnarray}
	Last inequality in Eq.(\ref{amproof}) is obtained by applying the following result from \cite{verst} on each of $\rho_1,\rho_2,...,\rho_n$:\\
	\textit{Result:} For any two-qubit state $\rho$(Eq.(\ref{st41})) with correlation tensor $W$$=$$\textmd{diag}(E_{1},E_{2},E_{3})$ such that $E_1$$\geq$$E_2$$\geq$$E_3$ along with concurrence $\mathcal{C},$ maximal Bell-CHSH violation is upper bounded by:
	\begin{equation}
		\sqrt{E_{1}^{2}+E_{2}^2}\leq \sqrt{1+\mathcal{C}^2}.
	\end{equation}
	Eq.(\ref{amproof}) corresponds to the bound given in Theorem.\ref{theo4}.
	This completes the proof.
 \section*{Appendix.E}\label{appnc}
 Let each of $S_i$ prepare an arbitrary two qubit entangled state $\rho_{A_i B}$ with concurrence $\mathcal{C}_i$. Let each $A_i$ perform a general single qubit measurement and similarly for Bob. We can define these measurements as : For Each $A_i$ $\rightarrow$ $\vec{a_x^i} \cdot \vec{\sigma}$ and for $B$ $\rightarrow$ $\bigotimes_{i=1}^n B_y^i = $ $\bigotimes_{i=1}^n \vec{b_y^i} \cdot \vec{\sigma}$. \\
 Given the set of local measurements described above, the expression $N_{Star}$ can be written as 
 \begin{eqnarray}\label{appE1}
		N_{Star} =&& {\frac{1}{2}|\Pi_{i=1}^n \textmd{Tr}((\vec{a_0^i}+\vec{a_1^i})\cdot \vec{\sigma} \bigotimes \vec{b_0^i}\cdot \sigma )\cdot \rho_{A_i B}|}^{\frac{1}{n}} +\nonumber\\&& {\frac{1}{2}|\Pi_{i=1}^n tr((\vec{a_0^i}-\vec{a_1^i})\cdot \vec{\sigma} \bigotimes \vec{b_1^i}\cdot \sigma )\cdot \rho_{A_i B}|}^{\frac{1}{n}}
	\end{eqnarray}

		Using Eqs.(\ref{app2},\ref{app3},\ref{app4},\ref{app5},\ref{extr3}), it is clear that $\forall i,j$ upper bounds of each element of $W^{(ij)}$ is independent of $p_{ij}$ terms. Consequently, the maximization strategy used in \cite{And} is applicable here and the measurement settings maximizing Eq.(\ref{ineqs}) here are same as that used in \cite{And} Following that we get the result:
	\begin{eqnarray}\label{app5iv}
		B_{n-star}&=&\sqrt{\Pi_{i=1}^n E_{i1}^{\frac{2}{n}}+\Pi_{i=1}^n E_{i2}^{\frac{2}{n}}}\nonumber\\
		&\leq&\sqrt{1+\Pi_{i=1}^n\mathcal{C}_i^{\frac{2}{n}}}
	\end{eqnarray}
This proves the Theorem.\\
However, let us now understand why the same bound(Eq.(\ref{app5iv})) is not applicable when at least one of the states $\rho_1,\rho_2,...,\rho_n$ is not entangled.\\
Following the same strategy as that we have used in Case.2 in Appendix.A, we get:
\begin{eqnarray}\label{separable21}
	B_{n-star}&=&\sqrt{\Pi_{i=1}^n E_{i1}^{\frac{2}{n}}+\Pi_{i=1}^n E_{i2}^{\frac{2}{n}}}\nonumber\\
	&\leq&\sqrt{P_1^{\frac{2}{n}}.\Pi_{i=2}^n E_{i1}^{\frac{2}{n}}+P_2^{\frac{2}{n}}.\Pi_{i=2}^nE_{i2}^{\frac{2}{n}}}\nonumber\\
	&\leq&\sqrt{P_1^{\frac{2}{n}}+P_2^{\frac{2}{n}}}\,\textmd{as } E_{ij}\leq 1\nonumber\\
\end{eqnarray}
Now, as $\rho_1$ is separable, we have Eq.(\ref{separable}). Also each of $P_1,P_2$$\leq$$1.$ Hence, from the last inequality of Eq.(\ref{separable21}), it is not always true that $\sqrt{P_1^{\frac{2}{n}}+P_2^{\frac{2}{n}}}$$\leq$$1.$ For instance one may consider $P_1$$=$$\frac{1}{2}$ and $P_2$$=$$\frac{1}{4}.$ However the expression 
$\sqrt{1+\Pi_{i=1}^n\mathcal{C}_i^{\frac{2}{n}}}$$=$$1$ as $\mathcal{C}_1$$=$$0.$ So, $	B_{n-star}$ not always less than $\sqrt{1+\Pi_{i=1}^n\mathcal{C}_i^{\frac{2}{n}}}.$ This is true whenever at least one of the states used in the network is not entangled. Hence Theorem.\ref{theo5} holds only when all the states used in the network are entangled. \\
Note that last inequality in Eq.(\ref{separable21}) is same as Eq.(72) when $n$$=$$2.$ In that case as the situation corresponds to that of the bilocal bound($B_{n-linear}$ for $n$$=$$2$), the upper bound of $B_{n-star}$ is given by $\sqrt{1+\Pi_{i=1}^n\mathcal{C}_i^{\frac{2}{n}}}$ for $n$$=$$2$. 
\section*{Appendix.F}
Proof of Theorem.\ref{sepcount}: Given that $m$ out of $n$ states are separable where:
\begin{equation}\label{appf1}
	m\geq \left\lceil\frac{n}{2}\right\rceil.
\end{equation}
 W.L.O.G. , let $\rho_1,\rho_2,...,\rho_m$($m$$\leq$$n$) be separable states used in the $n$-local star network.\\
 In that case, the upper bound of $\mathcal{I}_{n-star}$ gives:
\begin{eqnarray}\label{f2}
	B_{n-star}&=&\sqrt{\Pi_{i=1}^n E_{i1}^{\frac{2}{n}}+\Pi_{i=1}^n E_{i2}^{\frac{2}{n}}}\nonumber\\
	&\leq&\sqrt{\Pi_{i=1}^m E_{i1}^{\frac{2}{n}}+\Pi_{i=1}^m E_{i2}^{\frac{2}{n}}}\, \,\textmd{\small{as}  } E_{ij}\leq 1\nonumber\\
	&=&\sqrt{\Pi_{i=1}^m E_{i1}^{\frac{2m}{m.n}}+\Pi_{i=1}^m E_{i2}^{\frac{2m}{m.n}}}\nonumber\\
	&\leq&\sqrt{\frac{\sum_{i=1}^m E_{i1}^{\frac{2m}{n}}+\sum_{i=1}^m E_{i2}^{\frac{2m}{n}}}{m}}\,\, \textmd{(By } A.M.\geq G.M).\nonumber\\
&=&\sqrt{\frac{\sum_{i=1}^m( E_{i1}\cdot E_{i1}^{\frac{2m}{n}-1}+ E_{i2}\cdot E_{i2}^{\frac{2m}{n}-1})}{m}}\nonumber\\
&\leq&\sqrt{\frac{\sum_{i=1}^m( E_{i1}\cdot E_{i1}^{\frac{2m}{n}-1}+ E_{i2}\cdot E_{i1}^{\frac{2m}{n}-1})}{m}}\,\, \textmd{\small{(By Eq.(\ref{appf1}) and} }\nonumber\\
&&\quad\quad E_{i1}\geq E_{i2}\forall i)\nonumber\\
&=&\sqrt{\frac{\sum_{i=1}^m\cdot E_{i1}^{\frac{2m}{n}-1}(E_{i1}+ E_{i2})}{m}}\nonumber\\
&\leq&\sqrt{\frac{m}{m}}\nonumber\\
&=&1
\end{eqnarray}
Last inequality in Eq.(\ref{f2}) holds as $\rho_1,...,\rho_m$ are separable and hence by separability criterion\cite{See} we have:
\begin{equation}
	E_{i1}+E_{i2}\leq 1\,\forall i=1,2,...,m
\end{equation}

\end{document}